
\font\eightrm=cmr8
\font\eightbf=cmbx8
\font\eightit=cmti8
\font\eightsl=cmsl8
\def\smalltype{\let\rm=\eightrm
\let\bf=\eightbf
\let\it=\eightit
\let\sl=\eightsl
\baselineskip=9.5pt minus
.75pt
	\rm}
\magnification=1200
\baselineskip=18pt
\def\RR{{{I\negthinspace\!R}}}

\def\TT{{{T\negthinspace\negthinspace\!T}}}
\def\ZZ{{{Z\negthinspace\negthinspace\!Z}}}
\def\QQ{{{0\negthinspace\negthinspace\negthinspace\!Q}}}
\def\NN{{{I\negthinspace\!N}}}

\null
\centerline {\bf { PIECEWISE LINEAR MODELS}}
\centerline {\bf {FOR THE}}
\centerline {\bf {QUASIPERIODIC TRANSITION TO CHAOS.}}
\bigskip
\centerline  {February 2, 1995}
\bigskip
\bigskip
\centerline
{David K. Campbell \footnote{$^1$}{Physics Department,
University of Illinois, 1110 W. Green St., Urbana, IL 61801.},
Roza Galeeva \footnote{$^2$}{UMPA, ENS, 46 Allee d'Italie 69364 Lyon Cedex 07,
France},
Charles Tresser
\footnote{$^3$}{I.B.M. Po Box 218, Yorktown Heights, NY 10598.}, and
David J. Uherka \footnote{$^4$}{Mathematics Department, University of
North Dakota, Grand Forks, ND 58202-8376.}}   \bigskip
\bigskip
\bigskip
\bigskip
\bigskip
\bigskip
\centerline {\bf {Abstract}}
\bigskip
\bigskip
\bigskip
We formulate and study analytically
and computationally two families of piecewise linear degree one circle maps.
These families offer the rare advantage of being non-trivial but
essentially solvable models for the phenomenon of mode-locking
and the quasi-periodic transition to chaos. For instance, for these
families, we obtain complete solutions to several questions still largely
unanswered for families of smooth circle maps. Our main results
describe (1) the sets of maps in these families having some prescribed
rotation interval; (2) the boundaries between zero and positive topological
entropy and between zero length and non-zero length rotation interval; and (3)
the structure and bifurcations of the attractors in one of these
families. We discuss the interpretation of these maps as
low-order spline approximations to the classic ``sine-circle'' map
and examine more generally the implications of our results for the case of
smooth circle maps. We also mention a possible connection to recent experiments
on models of a driven Josephson junction.

\filbreak
\bigskip
\noindent
{\bf 1. Introduction.}
\bigskip

The phenomenon of ``mode locking'', in which two (or more) coupled
nonlinear oscillators having in general irrationally related
(``incommensurate'') intrinsic frequencies lock into a periodic motion
involving a rational (``commensurate'') ratio of their actual
frequencies, has been observed and studied experimentally
at least since the time of Huyghens [Hu], who discovered that
two pendulum clocks mounted on a common wall tended to ``synchronize''
their motions, so that both their frequencies and phases became
``locked'' together. Physical systems ranging from convecting fluids
[SHL],[MSE] through nonlinear electrical conductors [MM, BBJ] to chemical
reactions [MS] have been shown experimentally to exhibit not
only synchronization (1:1 mode locking) but also mode locking into many other
rational ratios (``p:q mode locking,'' for general integers p and q). Further,
although one is generally interested in mode-locking in the {\it temporal}
behavior ({i.e.}, the dynamics) of coupled oscillators, analogous phenomena can
occur in the {\it spatial} structure of certain solids in which interactions
with two different length scales compete [Bak], [BBr], [BrB].

In the biological realm, mode locking is both widespread and of
considerable importance to life processes. Already in the early
stages of the modern development of nonlinear dynamics, the
human heart was modeled as a nonlinear dynamical system [VdPVdM],
and the problem of mode locking in the context of cardiac arrhythmias
was studied in one of the central rigorous works on
nonlinear dynamics (see [Ar2] for a report of this early work).
Since then mode locking has been shown to be nearly
ubiquitous in biological systems, occurring for instance
(in addition to cardiac arrhythmias [PG], [Gl], [GB]) in the firings of
neurons,
in animal gaits, and in the
coupling of breathing and locomotion (for a concise elementary
introduction and references, see [StSt], and for more detailed accounts and
references, see [St], [GM], [Wi]).
It is thus not surprising that attempts to develop a detailed, quantitative
understanding of mode locking -- and of the associated ``quasiperiodic
transition to chaos'' [FKS],[JBB1],[JBB2], [ORSS] -- have played a central role
in
modern nonlinear dynamics and chaos theory.
Many of the applications to physical and biological systems
have focused on what is perhaps the simplest model of mode locking: continuous
mappings
of the circle onto itself. First introduced by Poincar{\'e} in
his efforts to model the motion of trajectories on tori, orientation-preserving
homeomorphisms and diffeomorphisms of the
circle continue to attract the attention of mathematicians both
as interesting and challenging dynamical systems in their
own right and in
their original context as Poincar{\'e} maps
induced by non-singular
flow on the two-dimensional torus [De], [Ar], [He]. More
recently, families of
circle endomorphisms which are deformations of rotations have appeared
as approximate models for the transition to chaos (as defined, for instance,
as the transition from zero to
positive topological entropy) in flows in $3$-spaces where tori
supporting non-singular flows become wrinkled and then destroyed as
some parameters are varied (see, e.g., [CTA], [Boyl], [MaT] and references
therein).

The classic example of a circle endomorphism is the so-called {\it standard
circle} or
{\it sine-circle} map, defined by the two-parameter
family :
$$f_{a,b}:\theta \mapsto (\theta
+a+{b\over 2\pi}\cdot
sin(2\pi\theta))_1\,,$$
with $(a,\,b)\in [0,\,1[\times \RR^+$,
and where we have used the notation:
$$(z)_n\buildrel \rm def\over =
z\,\bmod\,n\,\,.$$
The two parameters in $f_{a,b}$ correspond to the non-linearity ($b$) and the
{\it bare} mean rotation speed ($a$), {\it i.e.}, the mean speed at which
the map winds around the circle in the absence of nonlinearity. In the
natural science context of coupled oscillators, the two parameters correspond
to the strength of the coupling (the forcing of the oscillators on
each other) and the ratio of the individual frequencies of the two
oscillators, respectively. Note that the mean rotation speed {\it emerges} as a
property of the map, and one can generally control only the bare mean rotation
speed.

More generally, one would like to understand the dynamics
of maps, organized for instance in two-parameter families of the form
$$g_{a,b}:\theta \mapsto (\theta +a+h_b(\theta))_1\,,$$ where $h_b$ is
a one parameter family of maps on the unit interval
for which $0$ and $1$ are fixed points (see [GTr] for a review).
Whenever  $h_b$ is a family of smooth
maps, the description of the family $g_{a,b}$ becomes a very difficult
problem, many aspects of which are still beyond our analytic grasp, despite
important recent progress (see, e.g., [Me], [EKT] and references
therein). Although computational simulations and heuristic
calculations can in many cases provide considerable insight, one naturally
seeks a more tractable class of models from which to draw
analytic guidance. In this regard, it is useful to recall our
experience with the extensively analyzed
one-parameter endomorphisms of the {\it interval}. Here, the paradigm
is the quadratic family, exemplified
by the familiar ``logistic map'',
$$
f_{r}: x \mapsto rx(1-x).
$$
Many of the results which are quite difficult in the context of
the logistic map can be fairly easily established if one considers
the case of {\it piecewise linear} maps: the tent map and the
trapezoidal map are the two simplest examples, but higher-order
piecewise linear maps (such as the ``house'' map [Ga2], [HC])
have also been studied.

Following these clues,
we will focus our present study to the case in which
$h_b$ is {\it piecewise linear}, so that our circle map
has what one might call a ``sawtooth'' form. We stress, of course,
that even for piecewise linear circle maps, as in the case of
maps on the interval, a number of
questions remain deep, difficult, and open. Moreover, in
the cases of both interval and circle maps, piecewise linear maps contain
``pathologies,'' in the sense that some aspects of their dynamics and
bifurcation
structure are different from what happens in the smooth case. Many of
these pathologies are intimately related to the (easy) solvability of
some questions. Nonetheless, as we shall establish,
one can readily gain nearly complete analytic insight into
the bifurcation structure of piecewise linear circle maps. Further,
this insight is such that many of our results admit rather easy proofs
(although not all of them were that easy to find), and methods are most
often elementary. Thus a second important feature of these
piecewise linear maps is that they provide excellent pedagogical
examples allowing non-specialists to gain insight into
what physicists would call a non-trivial, solvable model for the quasi-periodic
transition to chaos.
This has motivated us to try to keep the bulk of the paper
accessible to non-specialists (and, in particular, non-mathematicians).
That the paper is accessible does not mean
that it is easy reading, since for compactness and precision,
most of our article is couched in mathematical terminology. Hence
non-mathematicians will likely wish to refer to Appendix A, which
provides the essential definitions and background material, before commencing
the body of the article; let us also mention [MSt] as a recent
general reference on one-dimensional dynamics.
A third motivation for our study was the possibility that detailed examination
of the piecewise linear models might lead to novel conjectures or suggest
methods of
proof for
broader classes of maps, including the smooth case. Indeed, as we discuss
below, this possibility has already been realized, and several new
results on the smooth have been stimulated by results obtained
in this study. (see, {\it e.g.}, [EKT],[GMT]).

The present article is the sequel to the announcement made in [UTGC],
which was itself in turn motivated by an unpublished manuscript by two of us
(D.K.C. and D.J.U.), dealing mainly with invertible maps. In outline,
we have organized the remainder of the paper as follows (refer to
Appendix A for definitions of the few
technical terms used here). In $\S$2, we describe the two types of
piecewise linear maps we consider, the {\it tip} maps and
the {\it  plateau} maps ({\it cf} Figure 1). In $\S$3, we organize
these maps in parametrized families. We argue that both the
tip maps and the plateau maps are most naturally organized in {\it
three}-parameter
families, which we slice into {\it two}-parameter families for the ease of the
discussion, to permit clearer graphical illustrations, and for purposes
of comparison with the standard circle map. In $\S$4, we associate to
any map $f$ a pair of {\it monotone} maps useful in the study of $f$.
In $\S$5, we define subsets of the parameter spaces, subsets whose
descriptions amount to capturing some essential aspects of the
bifurcation structure.
In $\S$6, we give some general results, counter-parts of known
results for smooth families.
In $\S$7, we describe the bifurcation structure for invertible maps
and for monotone but non-invertible maps.
In $\S$8, we describe the sets of maps with a given rotation interval.
In $\S$9, we assemble all our main results on the topology and
geometry of the  boundaries of topological chaos (meaning positive
topological entropy) and rotational chaos (meaning the rotation interval
has non-zero length). These results rely on some specific properties of tip
maps
proved in $\S$10, including aspects of the dynamics of tip maps
and their bifurcation structure, primarily involving periodic attractors
and related questions. Finally, in $\S$11, we present some general
conclusions, open problems,
and a discussion of possible physical realizations of these
piecewise-linear maps. In Appendix A, we have assembled some definitions
and classic background material on
subjects including degree, lift, rotation number and rotation interval with
their
basic properties, Denjoy theory, and topological  entropy.
Appendix B also contains (some more advanced) background material for the main
text.
It describes stunted families (see also [BMT], [DGMT] and references
therein) and concludes with a new result for such families, which is
used in $\S$9.
Appendix C explains how to compute the boundaries of some of the
regions described in $\S$5.
Appendix D describes numerical computations about the likelihood of
irrational rotation number for invertible maps. These computations
provide support for a conjectured result for which we mistakenly thought
we had a proof in
[UTGC]: interestingly enough, the numerical evidence was
gathered only after we realized our proof was faulty.

Throughout the main text and the appendices, all fractions will be
written in lowest order terms, except where otherwise specified. The
theorems in the main text
carry a number and a name which is an abstract of their content:
they are original (to the best of our knowledge), except where otherwise
specified. Some theorems are strictly speaking new, but their proofs are
merely rewritten versions
of similar proofs for smooth maps: this is acknowledged on each
occasion . The reason we include these proofs is that the original
proofs are not only scattered throughout the literature but are also often
published
in papers substantially more technical than the present one. The theorems in
the
appendices carry an uppercase letter or the name of their author: they
are (well) known, unless we indicate otherwise. Most technical
words appear first (except in this introduction) in {\it italicized} form,
followed by their definition or
endowed with a subscript which refers to a section of an appendix
in which the definition is given.
Finally, we note that on occasion we will quote later theorems in the
proofs of earlier ones; this is done from continuity and clarity in
the presentation. The essence of the problem being nonlinear, one
could hardly expect a totally linear exposition to be adequate !


\vfill
\eject
\bigskip
\noindent
{\bf 2. The maps.}
\bigskip

We use the term {\it circle map} to denote for short, a continuous map
from the circle $\TT=\RR / \ZZ$ -- we use the standard notation
in which $\RR$ denotes the set of real numbers and $\ZZ$ the integers--
to itself which has {\it degree one}$_{A-
1}$. In this paper, we consider two specific classes of piecewise-linear,
degree-one
circle maps, whose typical graphs are shown in Figure 1. In this
figure, as well as often in later discussions, we
freely identify circle maps with maps on the unit interval with
periodic boundary conditions. Thus the derivative of a map can be
thought of in an elementary sense, without reference to the smooth
manifold structure of the circle.

The first class we shall study are the {\it tip maps}
(the sawtooth maps with sharp teeth), a terminology whose
motivation
will become apparent when we introduce the second class. The tip
maps are defined by the following piecewise linear forms:
$$ t_{\Omega ,S; \delta}(x)\,\,=\quad\left\{ \eqalign{(Sx +
\Omega)_1  \quad \qquad \qquad \qquad \qquad if \qquad\qquad\,\,\,\,0 \leq
x \leq {1-\delta\over 2}\,,\cr ( s(x-{1\over 2}) +{1\over 2} +
\Omega)_1\quad \qquad \qquad  if\qquad  \,\,\,\,{1-\delta\over 2} \leq x
\leq
{1+\delta\over 2}\,,\cr ( S(x-1) + \Omega + 1)_1 \quad \qquad \qquad if\qquad
\qquad
\,\,\,\, {1+\delta\over 2} \leq x
	\leq 1\,,\cr
}\right .$$
where the {\it small slope} $s$ is given by $s= {[1-S(1-\delta)]\over
\delta}<S$, and $S\geq1$ is the {\it large slope}. We shall denote by
$I_{\delta}$ the interval $[{1-\delta\over 2} ,{1+\delta\over 2}]$ where the
slope is $s$. Figures 1 a-c display graphs of tip maps for three sets of
parameter values. Notice that the tip maps depend on {\it three}
parameters: the {\it bare frequency} parameter, $\Omega$ (corresponding
to the parameter $a$ in the usual parametrization of the standard family),
the large slope parameter, $S$ (corresponding to the strength of the nonlinear
coupling $b$ in the standard family), and a parameter $\delta$, which
determines the width of the piece with slope $s$. The tip map (thought of
as being deformed from a rotation) becomes non-invertible when the small
slope crosses zero (from above); when this
small slope is exactly zero, the map is termed {\it critical} and the
value of the large slope at criticality is $S_c(\delta) = {1\over (1-
\delta)}$, so that $\delta={S_c(\delta)-1\over S_c(\delta)}$. Recall that
the standard circle map (and more generally a
smooth circle map) is critical when it is monotonic but has zero
derivative at (at least) one point. We shall denote by  $T_{\omega ,S;
\delta}$, with $(\omega)_1=\Omega$, a {\it lift}$_{A-1}$ of
$t_{\Omega ,S; \delta}$.
\bigskip
The second class we call the {\it plateau maps}. These maps are defined
by $$p_{\Omega ,U; \delta}(x)\,\,=\quad\left\{ \eqalign {(S_c x +
\Omega)_1
	\qquad \qquad \qquad if \qquad \qquad \quad \,\,\,\,0 \leq x
\leq {1-\delta\over 2} + {H\over S_c}\,,\cr ({1\over 2}+H +
\Omega)_1
\quad \qquad \qquad if\qquad \,\,\,\, {1-\delta\over 2} + {H\over
S_c} \leq x \leq {1\over 2} - {H\over S_c}\,,\cr
( S_c ({1\over 2}-x)+ {1\over 2} + \Omega )_1\quad \qquad  if\qquad \qquad
\,\,\,\,{1\over
2} {H\over S_c} \leq x
			\leq {1\over 2} +{H\over S_c}\,,
\cr
({1\over 2}-H + \Omega)_1\qquad \qquad  \quad if\qquad \,\,\,\,{1\over 2}
+{H\over S_c}
		\leq x \leq {1+\delta\over 2} - {H\over S_c}\,,
		\cr (S_c (x-1) + 1 + \Omega)_1 \quad \qquad if\quad \qquad \qquad
\,\,\,\,{1+\delta\over 2} -{H\over S_c}
		\leq x \leq 1 \,,\cr }\right .$$
for $U \geq S_c = S_c (\delta)$, with $H = U -S_c$, and by
$p_{\Omega , U; \delta} = t_{\Omega ,U; \delta}$ for $U \leq
S_c$.
The maximum value of $H$ is $H_{max}(\delta) = {S_c \delta\over 4}$.
Figures 1 a,b,d illustrate three members of this family.
Set $U_{max} = S_c + H_{max}$; the graph of $p_{\Omega ,U;\delta}$
for $S_c \leq U \leq U_{max}$ is obtained from the graph of
$p_{\Omega ,U_{max};\delta}$ by cutting with two horizontal lines.
We shall denote by  $P_{\omega ,U; \delta}$, with
$(\omega)_1=\Omega$, a lift of $p_{\Omega ,U; \delta}$.
\bigskip
\noindent
{\bf 3. Parameter spaces.}

\bigskip
We first choose the parameter space for the maps $t_{\Omega ,S;
\delta}$:

\noindent
- clearly $\Omega$ can be thought of as an element of $\TT$, or of
$[0,1)$,

\noindent
- since $s>1$ (and $s>S$) when $S<1$, $S$ can be chosen in
$[1,+\infty)$,
\noindent
- since we want to consider only continuous maps, we have to take
$\delta$ in $[0,1)$.

Notice then that $S$ has to be equal to $1$ when $\delta =0$, and
that $s=1$ when $S=1$, so that the value of $\delta $ is irrelevant when
$S=1$. It follows that the value $0$ of $\delta$ can be omitted. In formal
mathematical notation, we
introduce the equivalence relation  $\sim$ on $\TT\times
[1,+\infty)\times (0,1)$, defined by  $(\alpha,1,\delta _1)\sim
(\alpha,1,\delta _2)$ for all
$\delta _1$ and  $\delta _2$, and by equality when the second coordinate of
the triplet is other than $1$. Then the parameter space for the tip maps can
be described as     $${\cal P}'_t=(\TT\times \{1\}\times
\{0\}\sqcup\TT\times [1,+\infty)\times (0,1))/\sim\,,$$ where the symbol
$\sqcup$ stands for disjoint union. However, we find it more convenient to
forget the quotient operation, and consider the parameter space  $${\cal
P}_t=\TT\times \{1\}\times \{0\}\sqcup\TT\times [1,+\infty)\times
(0,1)\,.$$
We shall study the bifurcations and aspects of the
dynamics throughout ${\cal P}_t$, by splitting this $3$-dimensional space
into disjoint cylinders $\delta=constant$ (see Figure 2). Other $2$-
dimensional subspaces of ${\cal P}_t$ are worth noticing, and we now list
some of them:

\noindent
- the  {\it critical surface} $S=S_c(\delta)={1\over 1-\delta}$ is the
locus of parameter triples corresponding to critical maps. It is the
disjoint union of the {\it critical lines} in the cylinders
$\delta=constant$.

\noindent
- The surface $S={1\over 1-2 \delta}$, where $S=-s$: this surface is
everywhere supercritical, except at $S=1$, and the {\it topological
entropy}$_{A-4}$ there is trivially equal to $log(S)$.

\noindent
- The surface $\delta={1\over 2}$, to which the previous one is
asymptotic as $S\to \infty$, and was the subject of some numerical
studies reported in [YH]
and [UC]. As discussed in [UC], the case $\delta={1\over 2}$ arises naturally
for the two-slope case when one is considering
successive $n$-slope piecewise linear approximations -- {\it i.e.},
linear splines -- to the standard family of circle maps.

\noindent
- The surface $\Omega=0$ is studied in [AM2]: these authors reparametrize
the part of this surface  above its critical line by  $S$ and $\sigma=-
s$ (with our notations), and show among other things that the entropy
strictly increases with either $S$ or $\sigma$, the other parameter
being kept fixed. For related results, see for instance [MV], [Ga], [GaT],
[GMT] and references therein.

The parameter space ${\cal P}_t$ has universal cover
$${\bf P}_t\,=\,\RR\times [1,+\infty)\times (0,1)\,,$$
with fundamental domain
$$[0,1)\times [1,+\infty)\times (0,1)\,.$$
This fundamental domain is partitioned in slices
$[0,1)\times [1,+\infty)\times \lbrace \delta \rbrace$ for $\delta \in
(0,1)$, and we shall use such slices to draw several figures of this
paper. In each slice, the line $S=1$ is called the {\it rotation line}, and
as discussed previously, all rotation lines represent the same set of
maps, to wit the rigid rotations. ${\bf P}_t$ is also the parameter
space for the lifts $T_{\omega ,S; \delta}$ of tip maps. For any $\delta
\in [0,1)$, we shall denote by ${\bf F}_{t; \delta}$ the two-parameter
family of lifts $T_{\omega ,S; \delta}$.
The parameter space ${\cal P}_p$ for
the maps $p_{\Omega ,S; \delta}$ coincides with ${\cal P}_t$ under
and up to the critical surface. Above this surface, $U$ is limited to the
range $[S_c(\delta),S_c(\delta)+H_{max}(\delta )]$. Hence
$${\cal P}_p=\lbrace \,(\Omega ,U,\delta) \,|\,\Omega \in \TT,\,\delta
\in [0,1)\,,\,U\in [1,S_c(\delta)+H_{max}(\delta)] \rbrace \,.$$
As for ${\cal P}_t$, we shall study the bifurcations and aspects of the
dynamics throughout ${\cal P}_p$, by splitting this $3$-dimensional
space into disjoint cylinders $\delta=constant$. We did not find any
other $2$dimensional subspaces of ${\cal P}_p$ worth noticing, but we
remark that the one-dimensional subspaces corresponding to fixed
$\delta$ and $\Omega$ are {\it stunted families}$_{B}$ beyond the
critical point. The universal cover of ${\cal P}_p$ is $$\lbrace \,(\omega
,U,\delta) \,|\,\omega \in \RR,\,U\in
[1,S_c(\delta)+H_{max}(\delta)],\,\delta \in [0,1)\rbrace
\,,$$
with fundamental domain
$${\bf P}_p\,=\,\lbrace \,(\omega , U,\delta)\, |\,\omega \in [0,1),\,U\in
[1,S_c(\delta)+H_{max}(\delta)],\,\delta \in [0,1)\rbrace
\,.$$
This fundamental domain is partitioned in slices
$[0,1)\times [1,S_c(\delta)+H_{max}(\delta)]\times \lbrace \delta
\rbrace$ for $\delta \in [0,1)$ and, like for tip maps, we shall use such
slices
for drawing purposes. ${\bf P}_p$ is also the parameter space for the
lifts $P_{\omega ,U; \delta}$ of plateau maps. For any $\delta \in
[0,1)$, we shall denote by ${\bf F}_{p; \delta}$ the two-parameter
family of lifts $P_{\omega ,U; \delta}$. We have represented ${\cal
P}_t$ and ${\cal P}_p$ in Figures 2 and 3.
\bigskip
\noindent
{\bf 4. Monotone bounds.}

\bigskip
With the natural partial order on real maps (see Appendix A-2), the {\it
monotone bounds} of a real function $F$ are the monotone upper bound
$F^+$, defined as the smallest monotone function greater than or equal to
$F$, and  the monotone lower bound $F^-$, defined as the largest monotone
function smaller than or equal to $F$: see Appendix A-2 for a review on
monotone bounds. All monotone bounds for the tip and plateau maps above
or on the critical line
are critical maps of our families. More specifically, the circle maps
obtained by projecting the monotone bounds read:
$$t_{\Omega ,S; \delta}^+(x)= t_{ \Omega+{S-1-S\delta\over
2}, S; {S-1\over S}}((x-{S-1-S\delta\over
2S})_1)\,,$$ $$t_{\Omega ,S; \delta}^-(x)= t_{
\Omega-{S-1-S\delta\over 2}, S; {S-1\over
S}}((x+{S-1-S\delta\over 2S})_1)\,,$$ and
$$p_{\Omega ,U; \delta}^+(x)=t_{ \Omega+\delta
(U-{1\over
1-\delta}),{1\over 1-\delta}; \delta}((x-U(1-\delta)+1)_1)\,,$$
$$p_{\Omega ,U; \delta}^-(x)= t_{ \Omega+\delta (-U+{1\over
1-\delta}), {1\over 1-\delta}; \delta}((x+U(1-\delta)-
1)_1)\,.$$ \bigskip
\noindent
{\bf Remark.} For the families of maps we consider in this
paper, the critical maps, the monotone bounds above the
critical line, and the
$F_\omega$'s of Theorem B in Appendix A-2, all can be
reinterpreted using the following construction when the rotation
number is in $(0,1)$ (see Figure 4 and [Ve1-Ve2]).

\noindent
- Start with the unit square, and draw the first diagonal and the
lines $\Delta_0:\,\,y=Sx$ and $\Delta_1:\,\,y=S(x-1)+1$, for
any $S>1$.

\noindent
- Choose any $A\leq{1\over S}$, and draw the square $Q_A$
with corners $(A,A)$ and $(A+{S-1\over S},A+{S-1\over S})$.

\noindent
- The pieces of $\Delta_0$ and $\Delta_1$ in $Q_A$ can be
completed to the graph of a critical circle map, where the
circle has length ${S-1\over S}$.

\noindent
- Now, varying A monotonically, one gets a parametrization of
either the critical maps with slope $S$, one of the families of
monotone bounds described previously, or the family of the $F_\omega$'s of
Theorem B for some map with the given $S$.

\noindent
- Using the kneading theory of {\it nicely ordered}$_{A-2}$ periodic orbits
(see [STZ] and references therein for a recent review of this centuries old
subject [Be], [Ch], [Sm], [Ma], [MH]), one can check that the $A$-interval
corresponding to any rotation number of the form ${p\over q}$ has length
${(S-1)^2\over S(S^q-1)}$ (see [Ve3] for details).
\bigskip
\noindent
{\bf 5. Some special sets in function and parameter space}
\bigskip

Let $C^0(\RR)$ be the space of continuous degree one lifts endowed with
the sup norm. Following [Boyl] and [MaT1] (which both extended to degree
one circle maps some pieces of the theory developed in [Ar] for
homeomorphisms) for each real number $\omega$ we define two subsets
of $C^0(\RR)$ as follows, where $I(F)$ denotes the {\it rotation
interval}$_{A-2}$:
$${\cal A}_\omega=\lbrace F\in C^0(\RR)\,|\,\omega\in I(F) \rbrace\,,$$
and
$${\cal L}_\omega=\lbrace F\in C^0(\RR)\,|\,\lbrace \omega \rbrace =I(F)
\rbrace\,.$$
It is easy to prove that both sets are path-connected for any
$\omega\in\RR$. In words, ${\cal L}_\omega$ is the subset
in which there is a {\it unique} rotation number $\omega$, whereas
${\cal A}_\omega$ is the set in which $\omega$ is among the
(many, possibly a continuum of) rotation numbers.

For $\delta \in [0,1)$, let ${\bf F}_{t; \delta}$ and ${\bf F}_{p; \delta}$
be the two-parameter families of lifts corresponding respectively to tip
maps and to plateau maps (see $\S$3). For each real number $\omega$, and
$u\in
\lbrace t,\,p\rbrace$, we define
$$A_{\omega,u}={\cal A}_\omega\cap {\bf F}_{u,\delta}\,,$$
and
$$L_{\omega,u}={\cal L}_\omega\cap {\bf F}_{u,\delta}\,.$$
Hence the $A_{\omega,u}$'s and the $L_{\omega,u}$'s are subsets of two-
dimensional spaces.

To simplify the language and the notations, {\it we
shall identify sets of standard lifts with the corresponding regions in
parameter space}: the context should tell which space we mean (parameter
space or function space), when the distinction is relevant. For
statements where the subscript $u$ could be either $t$ or $p$, we
usually suppress it. Thus $A_{\omega}$ stands for ``$A_{\omega,t}$
or
$A_{\omega,p}$" and so on.

For the region $A_{\omega}$, we have a decomposition into a
disjoint union
 $$A_{\omega} = A_{\omega}^- \sqcup A_{\omega}^o \sqcup
A_{\omega}^+\,, $$ where ``--'' stands for the subcritical region in
which $F\in \
{\bf F}_{u,\delta}$ is strictly increasing; ``o'' stands for the critical line
for which $F\in {\bf F}_{u,\delta}$ is increasing but has zero derivative at
at least one point; and ``+'' stands for the supercritical region, where $F\in
{\bf F}_{u,\delta}$ is non-monotonic. Similarly, we can write
 $$L_{\omega} = L_{\omega}^- \sqcup L_{\omega}^o \sqcup
L_{\omega}^+\,$$
and  Theorem 3 will tell us that $L_{\omega}^+=\emptyset$ when
$\omega$ is an irrational number.
Defining $A^{\ominus} = A^o \cup A^-$ and
$A^{\oplus} = A^o \cup A^+$, we can also write
$$A_{\omega} = A_{\omega}^{\ominus} \sqcup  A_{\omega}^+ =
A_{\omega}^- \sqcup A_{\omega}^{\oplus}\,.$$
Similarly, with $L^{\ominus} = L^o \cup L^-$ and
$L^{\oplus} = L^o \cup L^+$, we will write
$$L_{\omega} = L_{\omega}^{\ominus} \sqcup  L_{\omega}^+ =
L_{\omega}^- \sqcup L_{\omega}^{\oplus}\,.$$
\bigskip
\noindent
{\bf 6. General results concerning the sets $A_{\omega}$ and $L_{\omega}$}
\bigskip

We begin with three general results concerning the ``shape'' and
the boundaries of the sets $A_{\omega}$ and $L_{\omega}$ in parameter space.
Since these
results hold not only for our piecewise linear families, but also
for the standard family (and with similar proofs, see [Boyl], [MaT1] and
[ORSS]), we shall be fairly brief. Non-expert readers wishing to follow
closely should consult the references and work through our arguments
in detail.
\bigskip
\noindent
{\bf Theorem 1: Connectedness of the $A_{\omega}$'s.}
{\sl $\forall \omega \in \RR,\,\,A_{\omega}$ is a connected and simply
connected subset of ${\bf F}_{\delta}$.}
\bigskip
\noindent
{\bf Proof of Theorem 1.} This follows directly from the continuity
of the bounds of the rotation interval, which in turn is a
direct consequence of the combination of Theorems A(ii) and
B(i), and of the monotonicity of these bounds as a function
of ${\omega}$, which comes from Corollary A' and Theorem
B(i) in Appendix A-2.

\rightline {({\bf Q.E.D.} Theorem 1.)}
\bigskip
\noindent
Let us say that a set of curves in $\RR\times \RR^+$ is an {\it $L$-set of
curves} if for some $K$, and for all curves in the set, the first coordinate
is a uniform Lipschitz function of the second one, with Lipschitz constant
$K$. Once such a $K$ is determined, we use the term {\it $L$-set of
curves with Lipschitz constant $K$}. (Recall that $g$ is a {\it Lipschitz
function with Lipschitz constant $K$} if for all $x$ and $y$, $|g(x)-
g(y)|\leq K|x-y|$.)
\bigskip
\noindent  {\bf Theorem 2: A uniform Lipschitz
property.} {\sl

\noindent
i) - The left and right boundaries, $A_{\omega}^l$ and
$A_{\omega}^r$, of the $A_{\omega}$'s in $\RR\times \RR^+$ form an $L$-
set of curves.

\noindent
ii) - The collection of the sets
$$B_{\omega}^l=\lim _{{\theta}\to{\omega}^+}A_{\theta}^l\,,$$
$$B_{\omega}^r=\lim _{{\theta}\to{\omega}^-}A_{\theta}^r\,,$$
form an $L$-set of curves.

\noindent
iii) - For any $\omega$
$$A_\omega ^l=\lim _{{\theta}\to\omega ^-} A_{\theta}^l\,,$$
$$A_\omega ^r=\lim _{{\theta}\to\omega ^+}
A_{\theta}^r\,.$$ - For $\omega$ irrational
$$A_\omega ^l=B_{\omega}^l\,,$$
$$A_\omega ^r=B_{\omega}^r\,.$$
}
\bigskip
\noindent
{\bf Proof of Theorem 2.} We prove i) first, and first consider the
tip case, which covers the plateau case below and up to the critical
line: this proof is exactly as in the standard family (see [Boyl], [EKT]).
In fact we prove statement i) only for  $A_{p\over q}^l$ since the remaining
cases can be treated in an analogous way.

\noindent
{\bf Claim.} {\sl If
$T_{\omega ,S; \delta}\in A_{p\over q}^l$, the vertical cone in
$\RR\times \RR ^+$, with vertex at $T_{\omega ,S; \delta}$, and
boundaries made by lines with slopes ${2\over 1-\delta}$ and $-
{2\over 1-\delta}$, contains $A_{p\over q}^l$.}

\noindent
{\bf Proof of the Claim.} A direct computation yields
$$\forall \epsilon >0\quad ,\,\forall \mu\in [0,{2\epsilon\over 1-
\delta}]\quad T_{ \omega -\epsilon , S\pm ({2\epsilon\over 1-\delta}-Claim
\mu); \delta} \leq T_{\omega ,S; \delta}\,,$$
and
$$\forall \epsilon >0\quad ,\,\forall \epsilon ' >0\quad ,\,\forall
\mu\in [0,{2\epsilon\over 1-\delta}]\quad T_{ \omega -\epsilon
-\epsilon ', S\pm
({2\epsilon\over 1-\delta}-\mu); \delta} \leq T_{\omega -\epsilon ',S;
\delta}\,.$$  Consequently, using the continuity of the rotation number of
$T_{\omega ,S; \delta}^+$ as a function of the parameters, and
its monotonicity as a function of $\omega$, we have
$$\rho (
T_{ \omega -\epsilon , S\pm ({2\epsilon\over 1-\delta}-\mu);
\delta}^+) \leq {p\over q}\,,$$
and $$\rho (
T_{ \omega -\epsilon -\epsilon ', S\pm ({2\epsilon\over 1-
\delta}-\mu); \delta}^+) < {p\over q}\,.$$
Similarly
$$\forall \epsilon >0\quad ,\,\forall \mu\in [0,{2\epsilon\over 1-
\delta}]\quad T_{\omega +\epsilon , S\pm ({2\epsilon\over 1-
\delta}-\mu); \delta} \geq T_{\omega ,S; \delta}\,,$$
implies
$$\rho (
T_{ \omega +\epsilon , S\pm ({2\epsilon\over 1-\delta}-\mu);
\delta}^+) \geq {p\over q}\,.$$
The claim then follows from the three inequalities on $\rho$.

\rightline {({\bf Q.E.D.} Claim.)}

The proof for the plateau case above the critical line is even easier:
from the formula given in $\S$ 4, it readily follows that for any
$\delta$, and above the critical line, all curves $A_\omega ^l$ and
$B_\omega ^l$ are
parallel straight lines with slope $-{1\over \delta}$ in the $(\omega ,U)$
plane, while the curves $A_\omega ^r$ and $B_\omega ^r$ are parallel
straight lines with slope ${1\over \delta}$, which means that these curves
above the critical line form an  $L$-set of curves with Lipschitz constant
$\delta$.

\noindent
Statements ii) and iii) follow from i) combined with Theorems A(ii) and B(i)
in Appendix A-2.

\rightline {({\bf Q.E.D.} Theorem 2.)}
\bigskip
\noindent
{\bf Theorem 3: The Nature of $L_{\omega}$ and $A_{\omega}$ for irrational
$\omega$.}
{\sl \noindent
With $\QQ$ indicating the set of rationals, we have

\noindent
-(i). $\forall \omega \in (\RR \setminus \QQ),\,\, L_{\omega} =
A_{\omega}^{\ominus}$
in ${\bf F}_{\delta}$,

\noindent
-(ii). $\forall \omega \in (\RR \setminus \QQ),\,\,L_{\omega}$ is
a Lipschitz curve joining $S = 1$ to $S = S_c$
in the parameter space.

\noindent
-(iii).  The intersection of $A_{\omega}$ with any horizontal line
above the critical line in ${\bf F}_{\delta}$, is an interval of
positive length.}
\bigskip
\noindent
{\bf Proof of Theorem 3.}

\noindent
-(i). This statement can be reformulated as the following

\noindent
{\bf Claim } {\sl Above the critical
line, one cannot have $\rho(F^-)\,=\,\rho(F^+)\notin \,\QQ$.}

The proof of this claim goes as for the standard family (see [BlF] and
[CGT]) and relies on classical results recalled in Appendix A-2: we first
remark that there exist distinct $C^2$ smooth lifts $F_0$ and $F_1$ such
that, $\forall x\,\in\,\RR,\,\,\,F^-
(x)\,\leq\,F_0(x)\,\leq\,F_1(x)\,\leq\,F^+(x)$. If the claim is false, by
Theorem A(i), $\rho(F_0)\,=\,\rho(F_1)\,=\,\rho(F^+)\notin \,\QQ$, so
that $F_0$ (and $F_1$) has a dense orbit by Denjoy Theorem. Hence the
claim follows from Theorem A(iii).

\noindent
-(ii) follows from Theorem A(i)-(ii) and Theorem 2.

\noindent
-(iii) follows from (i) and Theorems A(ii) and B(i).

\rightline {({\bf Q.E.D.} Theorem 3.)}
\bigskip
\noindent
{\bf 7. Results for $A_{\omega}$ and $L_{\omega}$ below and on the critical
line.}
\bigskip

Before reporting our technical results, let us provide a few introductory
comments to guide the reader.
Below and on the critical line, the $L_{\omega}$'s and the $A_{\omega}$'s
coincide, {\it i.e.} $L^\ominus _{\omega} = A^\ominus _{\omega}$. By the
uniqueness of the rotation number for {\it fomeomorphisms}$_{A-2}$, these
sets form a partition of the parameter space. The topology and geometry of
the $L^\ominus _{\omega}$'s with irrational $\omega$ is described in
Theorem 3(ii). For the rational case, the topology of
$L^\ominus_{p/q}$ is covered by Theorem 4; we relegate the explicit
computations
(essential for constructing figures to scale) to Appendix C. In the statement
of Theorem 4, a {\it node} means a point in $L^\ominus_{p/q}$ above the $S=1$
line of trivial rotations ({\it i.e.}, we must have $S>1$),
at which point the left and right boundaries of
$L^\ominus_{p/q}$ coincide, so that taking out this point would disconnect
$L^\ominus_{p/q}$. The question of how the rational and the irrational $L^-
_{\omega}$'s ``share'' the Lebesgue measure of the parameter space below the
critical line is still open: numerical results are reported in Appendix D (see
also the last remarks at the end of the present section). The description of
behavior on the critical line is in Theorem 5, which slightly improves on the
general
result of [Boyd] specialized to our case (see also [Ve3]).

\bigskip \noindent
{\bf Theorem 4: Topology of the $L_{p/q}^{\ominus}$:
the ``Sausage'' Structure.} {\sl In  ${\bf F}_{\delta}$, $\forall q \in \QQ,
L_{p/q}^{\ominus} =  A_{p/q}^{\ominus}$ has $\lceil \delta  q \rceil -1$
nodes, where $\lceil x \rceil$ is the smallest integer greater than or equal
to $x$. Hence, the interior of $L_{p/q}^{\ominus}$ has $\lceil \delta q \rceil
$ connected components, each of which is simply connected.} \bigskip
\noindent
{\bf Proof of Theorem 4.} The graph of the $q^{\rm th}$ iterate of any map
$g$ in $L^\ominus_{p/q}$ intersects the diagonal at the periodic points of
the map (see Appendix A-2). If the intersections are transversal,
intersections occur for the graph of the $q^{\rm th}$ iterate of any
map close enough to $g$; this leads to the finite widths of the
mode-locking intervals (Arnold tongues). The only way that any small (additive)
perturbation of $g$ can generate no intersection of the graph of its $q^{\rm
th}$
iterate with the diagonal is
if the $q^{\rm th}$ iterate of $g$ is the identity; hence nodes
can only occur in this case. It thus remains to count
how often such situations occur in $L^\ominus_{p/q}$.
\noindent
For $S=1$, all maps are rotations, so in $L_{p/q}$, the number of points of
the periodic orbit to which ${1-\delta\over 2}$ belongs, and which are in
$[{1-\delta\over 2},{1+\delta\over 2}]$, is $\lceil \delta q \rceil$. For
$S=S_c$, on the left boundary of  $L_{p/q}^o$, ${1-\delta\over 2}$ is the
only point of $[{1-\delta\over 2},{1+\delta\over 2}]$ in a periodic orbit. By
continuity, going down from $S_c$ to $S$ on the left boundary of
$L_{p/q}^{\ominus}$, where  ${1-\delta\over 2}$ is a point of the unique
periodic orbit, $\lceil \delta q \rceil -1$ points of the orbit of ${1-
\delta\over 2}$ have to get in  $[{1-\delta\over 2},{1+\delta\over 2}]$. Each
time one point crosses ${1+\delta\over 2}$, the $q^{\rm th}$ iterate of the
map $t_{\Omega ,S; \delta}$ is the identity map $Id$, hence, we get a node
of $L_{p/q}^{\ominus}$. It only remains to show that each crossing occurs
once. But if $m$ points of the orbit of ${1-\delta\over 2}$ belong to $[{1-
\delta\over 2},{1+\delta\over 2}]$, the equality
$$t^q_{\Omega ,S; \delta}=Id$$
implies
$$s^{m-1}S^{q+1-m}=1\,,$$
or
$$ {\lbrack {1-S(1-\delta)\over \delta } \rbrack}^{m-1}S^{q+1-m}=1\, ,$$
and elementary algebra shows this equation in $S$ has at most one
solution in $(1,{1\over 1-\delta})$.
 \rightline {({\bf Q.E.D.} Theorem 4.)}
\noindent
{\bf Remarks.}

\noindent
- The fact that powers of piecewise linear circle homeomorphisms
can be the identity, which is the core of the sausage shape of the $L^-
_{p/q}$'s has, been known for a long time: for a deep result
using this fact, see [He2]. However the Arnold sausages have not
previously been described in print in a quantitative way,
except in [YH], which considers without giving a proof the case of
$\delta={1\over 2}$ ( we thank Leon Glass for pointing out this
reference to us).

\noindent
- As indicated in the introduction, the {\it mode-} or {\it frequency-locking}
phenomenon, first described (in the context of {\it phase-locking} or
{\it synchronization} in the 1:1 region) by Christian
Huyghens in the context of pendulum clocks hanging on the same wall, refers to
the fact that generically,  coupled non-linear oscillators beating at
rationally related frequencies will continue to do so, with the same
rational relation, under a small enough perturbation. For circle map
models of forced oscillators,
this corresponds to the non-zero widths of the $L^-_{p/q}$'s (or $A^-
_{p/q}$'s) whenever the non-linearity parameter is not zero ({\it i.e.},
away from the $S=1$ line of trivial rotations).
For generic smooth circle maps, and restricting to the real line of
entire nonlinear functions, no iterate can be the identity [He2], which
prevents any line in parameter space not intersecting the line of pure
rotations
from crossing the $L_{p/q}$'s at a single point and hence insures frequency
locking. Thus, by preventing frequency locking, the sausage shape
described in Theorem 4 can be viewed as a pathology of the piecewise-linear
models: typically, one rather tries to ensure frequency locking in physical
systems and assumes it in mathematical analyses. For
instance, in [Boyl], it is assumed that, except for rotations, no iterate of
the maps considered can be the identity. In the concluding section, we shall
discuss one physical system in which the ``Arnold sausage''
structure has been observed.
\bigskip
\noindent
{\bf Theorem 5: Likelihood of irrational rotation numbers for $S = S_c$.}
{\sl For any $\delta\in (0,1)$, the set
$E_{\delta}=(\Omega|\rho(t_{\Omega ,S_c(\delta);\delta})\not\in \QQ )$
has zero box dimension $d_B(\delta)$. }
\bigskip
\noindent
{\bf Remark.} It was previously known that the set $E_{\delta}$ has zero
Lebesgue measure and zero Hausdorff dimension ([Boyd], [Ve3]). For
generic families of smooth enough maps with at most two critical
points, such as the standard family, it is known that irrational rotation
numbers correspond to zero Lebesgue measure [Sw], and in fact form a
set of Hausdorff dimension smaller than $1$ on the critical line [Kh].
Furthermore, there is numerical
evidence that this dimension is positive and universal (with the
degeneracy ({\it i.e.}, the order) of the critical point as a modulus) [JBB1],
[JBB2].

\bigskip
\noindent
{\bf Proof of Theorem 5.} Fix $\delta\in (0,1)$. For $r>0$, let $K_r(\delta)$
be the number of $L^o_{p/q}$'s of length $\geq r$. Since we know from
[Boyd] that the set $E_{\delta}$ has zero Lebesgue measure, we get from
[Tri] that the
box dimension $d_B(\delta)$ of $E_{\delta}$ is:
$$d_B(\delta)=\limsup_{r\to 0}{ \log K_r(\delta)\over \log (1/r)}\,.$$
{}From the Remark in $\S$4
$$|L^o_{p/q}|={(S_c(\delta)-1)^2\over S_c(\delta)((S_c(\delta))^q-1)}\,,$$
where one only considers the  $L^o_{p/q}$'s with $q>1$, hence
 $$d_B(\delta)=\limsup_{r\to 0}{ \log K_r(\delta)\over \log
(1/r)}=\limsup_{q\to \infty}{ \log K_{|L^o_{p/q}|}(\delta)\over \log \lbrack
{S_c(\delta)((S_c(\delta))^q-1)\over(S_c(\delta)-1)^2}\rbrack}=0\,,$$
since $S_c>1$ and $K_{|L^o_{p/q}|}$ grows as a polynomial in $q$.

\rightline {({\bf Q.E.D.} Theorem 5.)}
\bigskip
\noindent
{\bf Remarks.}

\noindent
-  Using the above formula for $|L^o_{p/q}|$ and the well known
identity [HW]
$$\sum_{k=1}^{\infty}{x^k\over 1-x^k}\phi (k)={x\over (1-x)^2}\,,
$$
where $\phi(n)$ is Euler's totient function, which counts the integers
smaller than $n$ and coprime with $n$, it is easy to check directly that
$$\sum_{q=2}^{\infty}|L^o_{p/q}|={1\over S_c}\,,$$
so that by the Remark in $\S$4, the set $E_{\delta}$ has  zero Lebesgue
measure: this is essentially the proof given in [Ve3]. The proof given in
[Boyd] covers a much wider class of maps.

\noindent
- It was asserted in [UTGC] that $\forall S < S_c$ and $\forall
\delta \in (0,1), \mu  (\Omega| \rho(t_{\Omega ,S;\delta})\not\in \QQ ) >
0$, where $\mu (E)$ is the Lebesgue measure of the set $E$. However, the Lemma
(not only its proof) we thought we had to prove this measure property is
wrong.
On the other hand we have some numerical results and remarks
suggesting that the measure property is true (see Appendix D), in
contrast to a rigorous result in [VK] establishing zero
measure of $E$ for piecewise
smooth circle maps with a {\it single} singular point; the
resolution of this seeming contradiction is simply that our tip maps have two
singular points.

\noindent
- It is also worth noting that, for each $\delta \in
(0,1)$, there are countably many rational curves $l_{(n,m)}$, with $(n,m)\in
\ZZ ^+\times \NN$, in the $(\omega ,S)$ parameter space, which cross each
$L_{p/q}$ at at most one point. Clearly, irrational rotation numbers
correspond to a set of full measure on these curves. With $T_{ \omega
,S; \delta}$ standing for the lift of $t_{\Omega ,S; \delta}$ such that
$T_{\omega ,S; \delta}(0)\in [0,1)$, the curve  $l_{(n,m)}$ is the
solution of  $$T_{\omega ,S; \delta}^n({1-\delta
\over2})={1+\delta\over2}+m$$
Some of these curves were drawn in [YH] in the case when $\delta={1\over
2}$. In our Appendix C, two such curves are drawn for the case
of  $\delta= 0.6$ (see Fig. C16).

\noindent
- The measure property we conjecture to be true for our piecewise linear
maps is known to hold for the standard family, as a particular case of a
more general and quite deep result of Michael Herman [He1].
\bigskip
\noindent
{\bf 8. Maps with given rotation interval.}
\bigskip

The main result in this section is Theorem 8 which describes the sets of
maps with given rotation interval. The other Theorems describe
structure properties of maps satisfying certain conditions, which
are useful to prove Theorem 8 or the results in the next section
about the boundary of chaos. Some of the structure properties are
related to {\it nicely
ordered} orbits and related concepts, as reviewed in Appendix A-2. The
reader is reminded that ${\bf F}_{\delta}$ denotes either
of the tip or plateau map families (see $\S$5).
\bigskip
\noindent  {\bf Theorem 6: Uniqueness of the pair of well-ordered
periodic orbits in  $A^+_{p/q}$.}
{\sl

\noindent
- Any $F_{\omega ,T; \delta} \in {\bf F}_{\delta}$ in the interior of
$A^+_{p/q}$ has exactly two $p/q$-ordered orbits.

\noindent
- Exactly one of these orbits, denoted  ${\bf O'}_{p\over
q}$, is contained in the regions where $F_{\omega ,T;
\delta}$ is increasing.

\noindent
- There exists  $T_c^{\prime} (\omega ,\,{\bf F}_{\delta})>
S_c$ such  that for each $S < T_c^{\prime}(\omega ,\,{\bf
F}_{\delta})$, ${\bf O'}_{p\over q}$ is  unstable and the
other $p/q$-ordered orbit is stable.

\noindent
- There is a single $p/q$-ordered orbit, also
denoted  ${\bf O'}_{p\over q}$ for $F_{\omega ,T; \delta}$
on the
boundary of  $A^+_{p/q}$: this orbit contains the lifts of at least one
of the singular points of  $f_{\Omega ,T; \delta}$ with
$\Omega=(\omega )_1$. }  \bigskip
\noindent
{\bf Proof of Theorem 6.}

Using, e.g., the Remark in $\S$4 or the fact that $S>1$, the proof is an
exercise left to the studious reader; as a hint, consider
the  lift of the $q^{\rm th}$ iterate of the map zigzagging back and
forth across the line $y=x$.

\rightline
{({\bf Q.E.D.} Theorem 6.)}

\bigskip
Let ${\bf O}_{P\over Q}=\lbrace P_0,\,P_1,\dots,P_{Q-1}\rbrace$, be the
projection of ${\bf O'}_{p\over q}$ onto the circle, where we have set
$f_{\Omega ,T; \delta} (P_j) = P_{(j+1)_Q}$, and ${P\over Q}=({p\over
q})_1$. It follows from the properties of ${\bf O}_{P\over Q}$ that the two
critical points $C$ and $K$ of $f_{\Omega ,T; \delta} $ are in an arc $A$
bounded by two successive points $P_j$ and $P_k$ of ${\bf O}_{P\over Q}$,
which coincide when $Q=1$. Let $P'_j$ be a lift of $P_j$, $P'_k$ be the lift
of $P_k$ immediately to the right of $P'_j$, and let $C'$ and $K'$ be the
lifts of $C$ and $K$ in $[P'_j,\,P'_k]$. Let $F_{\omega ,T; \delta}$ be the
lift of $f_{\Omega ,T; \delta} $ such that $P'_j$ (and $P'_k$) has rotation
number ${p\over q}$
under $F_{\omega ,T; \delta}$, {\it i.e.},
$\underline{\rho}_{F_{\omega ,T; \delta}
}(P'_j)=\overline{\rho}_{F_{\omega ,T; \delta} }(P'_j)={p\over q}$. We
have
\bigskip
\noindent
{\bf Theorem 7: Boundary behavior.} {\it Let $F_{\omega ,T; \delta}\in
A^\oplus_{p\over q}$ and $I(F_{\omega ,T; \delta})=[\alpha,\,\beta]$

\noindent
i) ${p\over q}=\beta$ if and only if
$$F_{\omega ,T; \delta}(C')\leq F_{\omega ,T; \delta}(P'_k)\,,$$
\noindent
ii) ${p\over q}<\beta$ if and only if
$$F_{\omega ,T; \delta}(C')> F_{\omega ,T; \delta}(P'_k)\,,$$
\noindent
iii) ${p\over q}>\alpha$ if and only if
$$F_{\omega ,T; \delta}(K')< F_{\omega ,T; \delta}(P'_j)\,,$$
\noindent
iv) ${p\over q}=\alpha$ if and only if
$$F_{\omega ,T; \delta}(K')\geq F_{\omega ,T; \delta}(P'_j)\,,$$
\noindent
v) ${p\over q}\in Interior(I(F_{\omega ,T; \delta}))$ if and only if
$$F_{\omega ,T; \delta}(C')>F_{\omega ,T; \delta}(P'_k)\quad and
\quad F_{\omega ,T; \delta}(K')<F_{\omega ,T;
\delta}(P'_j)\,,$$
\noindent
vi) $\lbrace {p\over q}\rbrace =I(F_{\omega ,T; \delta})$ if and only if
$$F_{\omega ,T; \delta}(C')\leq F_{\omega ,T; \delta}(P'_k)\quad and
\quad F_{\omega ,T; \delta}(K')\geq F_{\omega ,T; \delta}(P'_j)\,.$$}
\bigskip
\noindent
{\bf Proof of Theorem 7.}
Statement i) follows readily from Theorem 6
and Theorem B: for ${p\over q}$ to be in the  interior of $I(F_{\omega ,T;
\delta})$, it is necessary and sufficient that  $\rho(F^-_{\omega ,T;
\delta})<{p\over q}$ and $\rho(F^+_{\omega ,T; \delta})>{p\over q}$.
Statement ii) is proven similarly, as are iii) and iv). Statements v) and
vi) follow from i) through iv).

\rightline{({\bf Q.E.D.}   Theorem 7.)}
\bigskip
\noindent
{\bf Theorem 8: Maps with given rotation interval.} {\it For each $\delta$
and each closed interval $I$, the set of lifts in  ${\bf F}_\delta$
with rotation interval $I$ corresponds to a connected and simply connected
subset $R_I$ of the parameter space $\RR\times \RR^+$ or $\RR\times
[0,H_{max}(\delta)]$. More precisely, there exists $K$ as given in the proof
of Theorem 2 such that:

\noindent
- If one endpoint of $I$ is irrational, the second boundary being the same
irrational number or a rational number, $R_I$ is an arc where the first
coordinate is a Lipschitz function, with Lipschitz constant $K$, of the
second one.

\noindent
- If the endpoints of $I$ are distinct irrational numbers, $R_I$ is a point.

\noindent
- When  $I=\lbrace {p\over q}\rbrace$, $R_I$ is bounded by two arcs
where the first coordinate is a Lipschitz function, with Lipschitz constant
$K$, of the second one; these two arcs intersect at a single point above the
critical line and on the rotation line, the other intersections being
described in Theorem 4.

\noindent
- If the endpoints of $I$ are distinct rational numbers, $R_I$ is above the
critical line and is bounded by two arcs where the first coordinate is a
Lipschitz function, with Lipschitz constant $K$, of the second one. These
two arcs have common endpoints and disjoint interiors.
}
\bigskip
\noindent
{\bf Corollary 8': Topology of $L_{p/q}^{\oplus}$.} {\sl
$\forall p/q \in \QQ, L_{p/q}^{\oplus}$ is a connected and simply connected
subset of ${\bf F}_\delta$. }
\bigskip
\noindent
{\bf Remark.} At the time [UTGC] was written, Corollary 8' was
conjectured to hold for reasonable smooth families (such as the standard
family) (see [Boyl] p. 378 and Figure 13, and [MaT1] p.213). This has since
been proved, together with most of Theorem 8, for the standard family: in
this family,
only the case of a rotation interval with distinct irrational end points
is open at this writing: the corresponding region is known to be
connected, and conjectured to be a point [EKT].
\bigskip
\noindent {\bf Proof of Theorem 8.} From Theorems 2, 4 and 7, it is
sufficient
to prove that each  $A_\omega ^l$ or $B_\omega ^l$ intersects each
$A_\theta ^r$ and $B_\theta ^r$ for $\theta<\omega$, and $B_\omega
^r$ at a unique point above the rotation line when $\omega$ and
$\theta$ are both rational or both irrational.
\noindent
In the case of plateau maps, the  existence and uniqueness of such
intersections follow readily from the proof of Theorem 2 in the
plateau case.
\noindent
In the case of tip maps, tongues boundaries above the critical line
no longer necessarily correspond to monotonic graphs when the
second coordinate is expressed as a function of the first (see
Figure 5). Hence we have to prove both existence and uniqueness of
the crossings.
\noindent
For any $\omega$, when $S$ is
large enough, the thinnest strip with sides parallel to the first
diagonal and containing the graph of $T_{\omega ,S; \delta}$, is as
wide as one wants. Hence:
\bigskip
\noindent
{\bf Lemma 8''.} {\it For any $\omega$, if $S$ is
large enough, $\omega$ is contained in the interior of $I(T_{\omega
,S; \delta})$.}
\bigskip
\noindent
Lemma 8'' takes care of the existence of the crossings and it remains
to control the intersection of $C_{\omega } ^l$ and  ${C'}^r_{\theta} $ ,
with $C$ and $C'$ standing throughout the remainder of the proof for
either $A$ or $B$.
\noindent
For that, we first recall that, by Theorem 2, the curves whose
intersections we consider are graphs of maps from the second coordinate
to the first one. Hence, in case of multiple crossings, we would get two
maps both with two slopes, say $(S,s)$ and $(S',s')$ with $S>S'>0>s'>s$ (or
$S'>S>0>s>s'$) both at the intersection of two curves.
\noindent
{}From kneading theory [MTh], [AM1], we can deduce that these two maps
would have the same kneading sequences, hence that the turning points
would have identical inverse legal paths (in the sense of [MSS]). In the
case when $\omega$ and $\theta$ are  both rational (see the rational case
below) or both irrational (see the irrational case below), we will show
that the intersection of $C_{\omega } ^l$ and  ${C'}^r_{\theta} $ is a unique
point. It follows, using Theorem 2-iii), that the intersection of
$C_{\lambda} ^l$ and  ${C'}^r_{\mu}$ is always a point or an arc. The
uniqueness
results we can obtain and the limit connectedness of the intersection in
all cases is enough to prove the theorem in the tip case. The missing
uniqueness results are conjectured to be true but, in the case when one
of $\lambda$ and $\mu$ is rational and the other irrational, our methods
work only when dealing with maps which have some iterate with slopes
everywhere greater than one in absolute value (such cases are treated like
the irrational case below).

\noindent
{\bf The rational case}. We isolate first the case when
$\omega={p\over q}$ and $\theta={p'\over q'}$ are rational numbers. Then
the map at the crossing of $C_{p\over q} ^l$ and  ${C'}^r_{p'\over q'}$ is a
Markov map {\it i.e.}, the orbits of the turning points are finite, and
together give a  Markov partition of the circle, whose set of boundaries is
forward invariant. The pieces of the partitions for both maps are labeled
similarly by kneading theory, as well as all inverses because of the
topological conjugacy. Since both absolute values of the slopes for one
map are greater than those for the second one, we see that all sufficiently
remote backward images of the initial partition for one map get
smaller than the similarly labeled pieces for the other map, which is
impossible since in both cases one has to cover exactly the same
circle.

\noindent
{\bf The irrational case.} While this question arises naturally
in the present context, the result is most easily understood by
referring to Theorem 16, which is proved independently
in a later section.

{}From Theorem 16, some iterate of
the maps at the crossing of $C_{\omega} ^l$ and  ${C'}^r_{\theta}$ have
slopes everywhere greater than one in absolute value, and by  Corollary
16' and kneading theory, two maps at the intersection of a given pair
of curves $C_{\omega} ^l$ and ${C'}^r_{\theta}$ are topologically
conjugate. We thus can use a specialization of the result in [GMT], that
two piecewise linear maps, one of which has an iterate with slopes
greater than one in absolute value, and one of which has all slopes in
absolute value greater than the other, cannot be topologically
conjugate.

\rightline
{({\bf Q.E.D.} Theorem 8.)}
\bigskip
\noindent
{\bf Theorem 9: Maps in $L^+_{p\over q}$.} {\it Any map $f\in L^+
_{p\over q}$ permutes cyclically $q$ arcs with pairwise disjoint
interiors, the restriction of $f^q$ to any of these arcs is a bimodal map
which fixes the end points, all these maps are smoothly conjugate to
each other, and the set of end points of the arcs is the periodic orbit
${\bf O}_{P\over Q}$.}
\bigskip
\noindent
{\bf Proof of Theorem 9.}  This proof is a repetition of the one given
in [Boyl] and [MaT1] for some families of smooth maps. In the case
when $q=1$, the Theorem follows directly from Theorem 7, hence
we assume $q>1$
in the remainder of this proof. The orbit  ${\bf O}_{P\over Q}$
determines a partition in $q$ arcs of its complement in the circle. Let
$J_0,J_1,\dots ,J_{q-1}$ be the corresponding closed segments, and
$f_i$ (respectively $(f^q)_i$) be the restriction of $f$  (respectively
$f^q$) to $J_i$. We assume $J_0$ is bounded by $P_j$ and $P_k$, where
we use the notations of the paragraph before Theorem 7. With these
notations, $f_0$ has exactly two critical points (or turning intervals
for plateau maps), while the other $f_j$'s are linear maps.
We have
$$(f^q)_0=f_{q-1}\circ f_{q-2}\circ \dots \circ f_0\,,$$
$$(f^q)_1=f_{0}\circ f_{q-1}\circ \dots \circ f_1\,,$$
$$..............................$$
$$(f^q)_{q-1}=f_{q-2}\circ f_{q-3}\circ \dots \circ f_{q-1}\,.$$
This implies that all these maps are bimodal and fix the end points.
The conjugacy property comes from
$$(f^q)_{q-1}=(f_{q-1})^{-1}\circ (f^q)_{0}\circ f_{q-
1}\,,$$ $$(f^q)_{q-2}=(f_{q-2})^{-1}\circ (f^q)_{q-
1}\circ f_{q-2}\,,$$ $$..............................$$
$$(f^q)_{1}=(f_{1})^{-1}\circ (f^q)_{2}\circ f_{1}\,,$$
since $f_1$, $f_2$,...,$f_{q-1}$ are homeomorphisms.

\rightline {({\bf Q.E.D.}   Theorem 9.)}
\bigskip
\noindent
{\bf 9. Topology and geometry of the boundaries of chaos and
rotational chaos.}
\bigskip

Let $h(f)$ be the {\it topological entropy}$_{A-4}$ of $f$, and $F$ be some
lift
of $f$.

Consideration of the various definitions and results in Appendix
A allows one to conclude that if $I(F)$ is not a single point, then $h(f)>0$.
Explicitly, as we now explain, $f$ then has a {\it horseshoe}$_{A-4}$ and hence
has
positive topological entropy by Theorem D (see Appendix A-4).

Since by hypothesis $I(F)$ is not a single point, it contains two distinct
numbers, say $\omega$
and $\theta$. By Corollary B' in Appendix A-2, one can find $x$ with
rotation number $\omega$ and $y$ with rotation number $\theta$. Let
$J$ stand for any of the two arcs bounded by the projection $X$ of $x$
and the projection $Y$ of $y$: a high iterate of this arc covers the
circle, hence $J$, as many times as we like. We shall say that a map
$f$ has {\it rotational chaos} if $I(F)$ is not a single point for some
lift $F$ of $f$, and is {\sl topologically chaotic} (in short {\sl
chaotic}) if $h(f)>0$. Hence, by our remarks above, we
see that rotational chaos implies chaos. However, the
converse is {\it not} true, as some of the following results will
explain. At this stage, we want to recall that, to the contrary of a
claim often published in the physics literature, chaos, in any
reasonable sense (not necessarily defined as we do by the
positiveness of the topological entropy) is not ubiquitous as soon as one
crosses the critical line: there are chaotic maps arbitrarily close to
the critical line, but not arbitrarily close to all points on the critical
line.

For the next few theorems, we need to introduce some additional
definitions. Let  $A_{p/q}^{stab}$ be that part of the region
$A_{p/q}$ in which $f$ has a stable $q$-periodic point,
with similar notations for $L_{p/q}^{stab}$, and let $L_{p/q}^{dbl}$
be that part of $L_{p/q}$ in which $f$
has only periodic orbits with periods of the form $2^m\cdot q$,
for some $m\geq 0$.

\noindent {\bf Theorem 10: Region of zero topological entropy}.

\noindent
i) {\sl
In any family ${\bf F}_\delta$,
$h(f) = 0  \quad \Leftrightarrow \quad f \,\in\, \bigcup_{\omega}
L_{\omega}^{\ominus} \cup \bigcup_{p/q} L_{p/q}^{dbl}$

\noindent
ii) Furthermore, in ${\bf F}_{t;\delta}$,
$h(f) = 0 \quad  \Leftrightarrow \quad f \, \in \, \bigcup_{\omega}
L_{\omega}^{\ominus} \cup \bigcup_{p/q} L_{p/q}^{stab}    $.
}
\bigskip
\noindent
{\bf Proof of Theorem 10.}
Statement i) follows from the combination of Theorem 7 and
Theorem C in Appendix A-4. Statement ii) is by combining i) with
Theorem 16.

\rightline {({\bf Q.E.D.}   Theorem 10.)}

\bigskip
\noindent {\bf Theorem 11:  Topology and Geometry of the boundary
of the circularly regular region.}
{\sl
The boundary of the circularly regular region  ${\bf F^{c-reg}}$ of
${\bf F}_\delta$ defined by ${\bf F^{c-reg}} =
\bigcup_{\omega \in \RR} L_{\omega}$, is connected and locally
connected, and has finite length. }
\bigskip
\noindent
{\bf Remark.} The corresponding statements are known to be false
in the standard family [BJa, FT].
\bigskip
\noindent
{\bf Proof of Theorem 11.} In both cases, from Theorem 2, we only
need to
prove that the height of $L_{p/q}^+$ above the critical line is a
function of ${p\over q}$, with an upper bound going to 0
exponentially as $ q \rightarrow \infty$. Slightly more than that is
given by Lemma 11'. \bigskip
\noindent {\bf Lemma 11': Height of $L_{p/q}^+$ in
${\bf}F_{\delta}$.} {\sl
The height of $L_{p/q}^+$ above the critical line in
${\bf}F_{\delta}$, is a decreasing function of $q$, with an upper
bound going to 0 exponentially as $ q \rightarrow \infty$. }
\bigskip
\rightline {({\bf Q.E.D.}   Theorem 11.)}
\bigskip
\noindent {\bf Proof of Lemma 11'.} This amount to computations,
since, by Theorems 7 and 9, we know that (using the notations of
Theorem 9), at the upper tip of $L_{p/q}^+$,  $(f^q)_{i}$ is a map on an
interval which covers the interior of this interval three times, and
following the proof of Theorem 9 tells us more precisely what is this
map.

We first treat the case of ${\bf}F_{t;\delta}$. Then, the map $(f^q)_{i}$
is as represented in Figure 6-a, {\it i.e.}, is bimodal and fixes the end
points, with two increasing segments of slope $S^q$, and one
decreasing segment with slope $S^{q-1}\sigma$, where $\sigma=
|s|={S(1-\delta)-1\over \delta}$. Since each of the segments where the
map $(f^q)_{i}$ is monotone covers $J_i$, we have $$(|J_i|-{2|J_i|\over
S^q})S^{q-1}\sigma =|J_i|\,,$$ hence
$$(1-\delta)S^{q+1}-S^q+(\delta -2)S+2=0\,. \qquad(*)$$
It is easy to prove that (*) has a unique solution greater than
$S_c={1\over 1-\delta}$, which yields an alternate proof of Corollary
8'. What we seek here is an upper bound for this solution.

So let
$$g(x)=(1-\delta)x^{q+1}-x^q+(\delta -2)x+2\,, $$ $$g'(x)=(q+1)(1-
\delta)x^{q}-qx^{q-1}+(\delta -2)\,, $$ $$g''(x)=qx^{q-2}(x(q+1)(1-
\delta)+1-q)\,. $$
Notice that $g''>0$ when $x\geq S_c$, that $g(S_c)={-\delta \over 1-
\delta}<0$, and that $g'(S_c)={1\over (1-\delta)^q}+\delta-2$, hence
$g'(S_c)>0$ for $q$ large enough. One step of Newton's method with
$S_c$ as a starting point yields $$y_1={1\over 1-\delta}-{{-\delta
\over 1-\delta}\over {1+(\delta -2)(1-\delta)^{q-1} \over (1-
\delta)^{q-1}}}\,,$$ which, since $g''>0$, is an upper bound of the
solution of (*) greater than $S_c$, and we have $$y_1-S_c={(1-
\delta)^q\over (1-\delta)+(\delta-2)(1-\delta)^q}<2(1-\delta)^{q-
1}\,.$$

In the case of ${\bf}F_{p;\delta}$, the map $(f^q)_{i}$ is as represented
in Figure 6-b, from which we get the height $h$ of $L^+_{p/q}$ above the
critical line:
$$h={S_c-1\over 2(S_c^q-1)}\,,$$
hence
$$h<{(1-\delta)^{q-1}\over 2}\,.$$
This takes care of the estimates. The computation also shows that the
height of $L_{p/q}^+$, in both families, only depends on $q$, hence is
the same as the height of $L_{1/q}^+$. By [BT], this implies that the
height of $L_{p/q}^+$ is a decreasing function of $q$.

\rightline {({\bf Q.E.D.}   Lemma 11'.)}
\bigskip
The circularly regular region for the tip maps with $\delta =
{1\over 4}$ is illustrated in Figure 7. Only the contributions of
a few $L_{p/q}$'s (${p\over q}\in [0,{1\over 2}]$ with $q\leq
11$ ) are represented.
\bigskip
\noindent {\bf Theorem 12: Topology and Geometry of the boundary
of the topologically regular region.} {\sl The boundary of the
topologically regular region ${\bf F^{reg}}$ of ${\bf F}_\delta$ defined
by  $f \in {\bf F^{reg}}  \Leftrightarrow h(f) = 0$, is connected and
locally connected
and has finite length, except in the plateau case when $S$ is too
small, in which case, it is made of arcs of finite length with both end
points on the upper boundary $H=H_{max}$ of the parameter space.}
\bigskip
\noindent
{\bf Proof of Theorem 12.}
We first consider the case of tips maps. By Theorem 10, we have to
control the boundary of $L_{p/q}^{+,stab}$ in $L_{p/q}$. This boundary
is where $\sigma S^{q-1}=1$, hence $S$ is the greatest root of $(1-
\delta)x^q-x^{q-1}-\delta=0$ (hence is a strictly decreasing function
of q), and the boundary of $L_{p/q}^{+,stab}$ in $L_{p/q}$ is a segment
parallel to the $\omega$ axis. Since it is clear (from the proof of
Theorem 11) that the tip of $L_{p/q}$ is beyond the boundary of
$L_{p/q}^{+,stab}$, the estimate in the proof of Theorem 11 implies
the tip maps part of Theorem 12.

The case of the plateau maps is solved by the following Lemma,
together with the estimate in the proof of Theorem 12, since the tip
of  $L_{p/q}$ corresponds to a chaotic map.
\bigskip
\noindent
{\bf  Lemma 12': The boundary of topological chaos in $L_{p/q}$ for
plateau maps.}
{\sl

\noindent
- When going up from the critical line to the positive entropy region
inside $L_{p/q}$ in ${\bf F}_{p;\delta}$, one  must cross a complete
cascade of period doubling bifurcations.

\noindent
- The boundary of the region where
$h(p_{\Omega ,U;\delta})=0$ in $L^+_{p/q}$ is a connected and locally
connected curve, made of at most countably many smooth  pieces
crossing the gaps of a Cantor set imbedded in the two  dimensional
parameter space.

\noindent
- This boundary has finite length, bounded by ${1\over \delta}$ times
the width of $L_{p/q}$ on the critical line.

\noindent
- Maps corresponding to parameter values on this
curve have zero topological entropy}
\bigskip
\rightline {({\bf Q.E.D.}   Theorem 12.)}
\bigskip
\noindent
{\bf Proof of Lemma 12'.}
In Appendix B, we show that, for each $S>1$, the parameter space for
the
maps  $c_{S,\mu;a,b}$ is a 3-cell ${\bf C}_S$. Now remark that for any
plateau map $f=p_{\Omega ,U;\delta}$ in $L^+_{p/q}$, where $U=S+H$, the
$q^{\rm
th}$ iterate of $f$ restricted to any $I_i$ and renormalized to the
unit interval, is some map $c_{S^q,\mu;a,b}$. From the proof of
Theorem 2, we know that all $L^+_{p/q}$'s are exact copies of each
other, up to uniform dilatation, and a computation shows that, up to
a homothetie, $L^+_{p/q}$ is the surface ${\Pi}_{S^q}$ in ${\bf
C}_{S^q}$, as represented in Figure 8 according to whether $S^q\leq
3$, $3<S^q<5$, or $5\leq S^q$. It is easy to check that horizontal
lines in the $(\Omega ,S)$-plane are sent to horizontal lines in the
$(a ,b)$-plane, and that vertical lines in the $(\Omega , S)$-plane
are sent to vertical lines in the $(a ,b)$-plane, hence ${\Pi}_{S^q}$ is
a plane, and cuts  ${\bf C}_{S^q}$ on a triangle ${\bf T}_{S^q}$
symmetrical with respect to its height: we leave to the reader to
check that the equation of this plane reads $\mu={1\over
2S^q}[S^q+1+2(a-b)]$.

The boundary of chaos in each $\mu$-cut of ${\bf C}_{S^q}$, is
described by Theorem E, and consists in the closure of a set of lines
with slopes $1$ or $-1$. By Theorem F, the surface corresponding to
one of these lines as $\mu$ varies cuts ${\Pi}_{S^q}$  in a line whose
absolute value of the slope at each point is not greater than the slope
of the sides of the triangle. The homothetie of the triangle ${\bf
T}_{S^q}$ to $L^+_{p/q}$ gives it sides with ${1\over \delta}$ as
absolute value of the slope.
\rightline {({\bf Q.E.D.}  Lemma 12'.)}
\bigskip
The region of zero topological entropy for the tip maps with $\delta =
{1\over 4}$ is illustrated in Figure 9. Only the contributions of a
few $L_{p/q}^{stab}$'s (${p\over q}\in [0,{1\over 2}]$ with $q\leq 11$ )
are represented.
\bigskip
\noindent
{\bf Remarks.} As for Theorem 11, the statements corresponding to
Theorem 12 are known to be false in the standard family [FT]. We
conjecture that the boundary of chaos inside each the tongues of the
standard family is of finite length.

Figure 10 displays some bifurcation lines of the plateau family with
$\delta={1\over 2}$. The second remark after the proof of Lemma F' in
Appendix B explains some of the straight lines one observes in this
figure, at least the piece of these straight lines insides the
$L^+_{p/q}$'s. To understand these straight lines more globally, notice
that the condition for a plateau boundary to be part of a periodic orbit
with a given combinatorics corresponds to a linear relation between
$H$ and $\Omega$: this is because of the way $H$ and $\Omega$ appear in
the formula of the maps. The $S-H$ or $S-\Omega$ relations would be
more complicated. \bigskip
\noindent
{\bf Theorem 13: A stratification of the plateau  family.} {\sl
For each $c>0$, the region of the parameter space of
${\bf F}_{p;\delta}$ where $h(f)=c$, is connected and simply connected,
except when $S$ is too small, in which case, each connected component
contains two pieces of the upper boundary $H=H_{max}$ of the parameter
space. When $U$ is increased above the critical line in  $p_{\Omega
,U;\delta}$ (with $\Omega$ and $\delta$ kept fixed),  no periodic orbit is
destroyed.
}
\bigskip
\noindent
{\bf Proof of Theorem 13.}
The second statement is an easy consequence of the fact that varying $U$
above the critical line in  $p_{\Omega ,U;\delta}$ with $\Omega$ and
$\delta$ kept fixed, correspond to a stunted family. Similarly, the
topological entropy is a non-decreasing and continuous function of
$U$, with $\Omega$ and $\delta$ kept fixed. Since the entropy is
continuous in ${\bf F}_{p;\delta}$, the first statement follows
(see [Msz]).

\rightline{({\bf Q.E.D.}   Theorem 13.)}
\bigskip
\noindent
{\bf 10. Supercritical tip maps with stable behavior.}
\bigskip

We begin with a completely elementary ({\it i.e.}, both the
techniques and concepts are elementary) proof of the following
result, which can also be obtained along the lines of the proof of
Theorem 16.
\bigskip
\noindent
{\bf Theorem 14: Stability implies Well-Orderedness for periodic
orbits.} {\sl Any stable periodic orbit of a tip map is ${p\over q}$-
ordered for some ${p\over q}$. Any marginally stable periodic orbit
${\bf O}$ of a tip map $f$ is either ${p\over q}$-ordered for some
${p\over q}$, or $f$ has a  marginally stable periodic orbit ${\bf O'}$
which is ${p\over q}$-ordered for some ${p\over q}$, the product of
the slopes along the orbits is $-1$, and ${\bf O'}$ has period $2q$ and
rotation number ${p\over q}$.}
\bigskip
\noindent
{\bf Proof of  Theorem 14.} Only maps above the critical line need to
be considered, since the statement is obvious elsewhere (where only
the first case occurs in the marginally stable case). In this proof
$[\alpha ,\beta]$ stands for the arc delimited by $\alpha$ and
$\beta$, where the orientation is not specified as going
counterclockwise from $\alpha$ to $\beta$, but by the fact that
$[\alpha ,\beta]\subset I_\delta$, where $I_\delta$ is the arc going
counterclockwise from ${1-\delta \over 2}$ to ${1+\delta \over 2}$.

- We first consider the stable case.

Hence, assume some supercritical tip map $f$ has a stable periodic
orbit ${\bf O}$ with period $q$, which is not ${p\over q}$-ordered for
any $p$. We shall organize the proof by the number $m$ of points of
${\bf O}$ belonging to the interior of the arc $I_\delta$ (which is the
arc carrying the slope $s$). Notice that the absolute value of the
slope of $f^q$ at any point of ${\bf O}$ is given by $\sigma ^mS^{q-
m}$, where we set again $\sigma=|s|$, and stability means $\sigma
^mS^{q-m}<1$.

\noindent
- Since $S>1$ above the critical line, the case $m=0$ is solved
trivially.

\noindent
- The case $m=1$ can be reduced to the case $m>1$  by a simple
surgery method as follows. Let $x$ be the unique point of ${\bf O}$ in
$I_\delta$, $x_L$ and $x_R$ the neighbors of $x$ in  ${\bf O}$, and
$X$, $X_L$ and $X_R$  lifts of these points which are consecutive
among lifts of point in ${\bf O}$. Since ${\bf O}$ is not ${p'\over q'}$-
ordered for any ${p'\over q'}$, either $F(X_L)>F(X)$ or $F(X_R)<F(X)$
for any lift $F$ of $f$. It is then easy to construct a circle map $f'$
such that ${\bf O}$ is an orbit of $f'$, and the method described below
for $m\geq 2$ works for $f'$: the details are explained in Figure 11 in
the case $F(X_L)>F(X)$.

\noindent
- In the case $m>1$, let $a_1,a_2,\dots,a_m$ be the points of ${\bf
O}$ in $I_\delta$, ordered clockwise. Define $n(i)$ as the time the
orbit of $a_i$ spends out of $I_\delta$ before returning to
$I_\delta$, {\it i.e.}, $$f^{n(i)}(a_i)\in Interior(I_\delta),\,\,{\rm
and}\,\,f^{m}(a_i)\not\in Interior(I_\delta)\,\,{\rm for}
\,\,0<m<n(i)\,,$$
and notice that, for $n=\inf_i(n(i))$, we have  $\sigma S^n<1$.

Let $j$ stand for the smallest $i$ such that $n(i)=n$.

- If $j=1$, with $a_k=f^n(a_1)$, the image under $f^n$ of the arc
$[a_1,a_k]$ has to be strictly included in itself by $\sigma S^n<1$,
but this is impossible because there are exactly $k$ points of ${\bf
O}$ in this arc.

- The case $j=m$ is treated as the case $j=1$.

- If $j\not\in \lbrace 1,m\rbrace$, set $a_k=f(a_j)$.

\qquad- If $k=1$, it is plain that $f^n(a_{j-1})\in I_\delta$, which
contradicts the minimality of $j$.

\qquad- If $k=m$, argue like when $j=1$.

\qquad- If $k\not\in \lbrace 1,m\rbrace$,  $f^n(a_{j-1})$ has to be
out of $I_\delta$, otherwise, argue like when $k=1$. Since $\sigma
S^n<1$, the arc $[a_{j-1},a_j]$ has to be longer than the arc
$[a_j,a_m]$, which, by the same $\sigma S^n<1$, implies that
$f^n([a_{j},a_m])$ has to be included in $[a_{j-1},a_j]$, but this is not
possible since there is no point of ${\bf O}$ in this arc.

- We now consider the marginally stable case, {\it i.e.}, $\sigma ^mS^{q-
m}=1$. If $n(i)$ is not constant, we proceed as before. So assume that
$n(i)=n$ for
all $n$, and that ${\bf O}$ has period $q$, but is not ${p\over q}$-
ordered for any $p$. Then, $\sigma S^n=1$, so that if $f^n(A_s)=A_t$,
we also have $f^n(A_t)=A_s$, and $m=2$. The middle point of the arc
$[A_s,A_t]$ has to be periodic with period ${q\over2}$, and if it
would not be ${p'\over {q\over2}}$-ordered, we would get a
contradiction by modifying $f$ so that the orbit of this point is
preserved, but the orbit is stable.
\rightline {({\bf Q.E.D.}    Theorem 14.)}
\bigskip
The following result is not only completely elementary, but also quite
easy, the only tricky case being covered in the discussion for Theorem
4. The proof is left to the reader, with the special case of a stable
fixed point explained in Figure 12.
\bigskip
\noindent
{\bf Theorem 15: A corner bounds each basin.} {\sl Given a tip or
plateau map $f$, for each stable periodic orbit with period $p$, there
is a point $x$ in the orbit such that the shortest of the intervals
centered at $x$ and bounded by one of the corners ${1-\delta\over 2}$
and ${1+\delta\over 2}$, is mapped into itself by $f^p$. Hence one of
the corners is in the immediate basin of attraction of any stable
periodic orbit, and there are at most two stable periodic orbits. If $f$
is not supercritical, both corners are in the immediate basin of
attraction of any stable periodic orbit, but generally not in the same
connected component of the basin. Hence there is at most one stable
periodic orbit. If stable is replaced by marginally stable, and the
slope is one, $f^q$ is a rotation. If stable is replaced by marginally
stable, and the slope is $-1$, there is a periodic interval, and all
points in it, except for the original periodic orbit, have period $2q$.}
\bigskip
\noindent
{\bf Remark.} It follows readily from Theorem 15 that there are at
most 2 stable periodic orbits for any map in the family, a statement
also known
to be true for the standard family, and easy to prove in the plateau case.
The general organization of the stability and bistability regions in the
parameter space of the standard family is closer to what we get for our
plateau maps (conjecturally, the differences between the two cases are
metric, but not topological except for the existence of behavior in the
plateau case not matched in the standard family inside the $L_{p/q}$'s, and
what is covered by
Theorems 11 and 12), than to the peculiar structure in the tip cases,
to be described in Theorem 17. In the subcritical case, bounds on the
number of periodic orbits for real analytic maps beyond the case of
the standard family are given in [He3].  \bigskip
\noindent
{\bf Theorem 16:  When there is no stable nicely ordered periodic orbit,
some iterate has all slopes out of [-1,1].} {\sl Let $f$ be a supercritical tip
map with no stable or marginally stable, nicely ordered periodic orbit.
Then for some $n$, the absolute value of the slopes of $f^n$ are all
greater than one.}
\bigskip
\noindent
{\bf Remark.} Theorems 14 and 16 can be reinterpreted as giving new
characterizations of nice order, a question which has been the object
of many investigations since the 18$^{\rm th}$ century (see [Be],
[Ch],
[Sm], [Ma], [MH], as well as [STZ] and references therein).
\bigskip
\noindent
{\bf Proof of  Theorem 16.} Assume Theorem 16 is false. Then there
exists some $x\in I_{\delta}$ and a smallest $n>0$ such that
$f^{n+1}(x)\in I_{\delta}$. It is then clear that $sS^n\leq 1$. Let
$I_{x}$ be the maximal interval containing $x$ and such that
$f^{n+1}(y)\in I_{\delta}$ for all $y\in I_x$. Since $f^{n+1}$ is
orientation reversing and either contacting or isometric on $I_x$, it
is easy to deduce that $I_x$ must contain an end point of $I_{\delta}$
and to further deduce that $I_x$ must contain a fixed point of
$f^{n+1}$, hence a stable periodic orbit for $f$. The rest follows from
Theorem 14.
\rightline {({\bf Q.E.D.}    Theorem 16.)}
\bigskip
\noindent
{\bf Remark.} The proof given here of Theorem 16 is quite specific to the
tip maps, which makes this theorem quite remarkable: one would more
generally like to know whether a piecewise linear map which is not a
homeomorphism and which does not have a stable periodic orbit must
have an iterate with all slopes greater than one in absolute value. In
this context see the results and the discussion in [MTr].
\bigskip
\noindent
{\bf Corollary 16': Density of the preimages of the turning points when
no stable nicely ordered orbit exists.} {\sl For any tip map with no stable
nicely ordered periodic orbit, the set of all preimages of the corners
${1-\delta\over 2}$ and ${1+\delta\over 2}$ is dense in the circle.}
\bigskip
For the next theorems, we define the {\sl stability region} ${\bf
R}_{\delta}$ of ${\bf F}_{t; \delta}$ as the set of $t_{\Omega ,S;
\delta}$'s with $0\leq\Omega<1$ in $\bigcup_{p/q} interior
(A_{p/q}^{\oplus,stab})$. The way the stability region is placed with
respect to the $A_{p/q}$'s is illustrated in Figure 13 for $\delta
={1\over 4}$, and global views of ${\bf R}_{\delta}$ are displayed in
Figure 14 for $\delta ={1\over 2}$ and $\delta ={3\over 4}$.  Until the
end of the main text, fractions are supposed to belong to the interval
$[0,1]$. The {\sl Farey sequence of order $i$} is the ordered set of
reduced  fractions in $[\,0,\,1]$, with denominators up to $i$ .  Two
fractions ${p\over q}$ and ${p'\over q'}$ in $[\,0,\,1]$  are {\sl Farey
neighbors} if they are consecutive in some Farey sequence. Let $N(i)$ be
the number of terms in the Farey sequence of order $i$ for $i > 0$, with
$N(0) = 1$, and let $N'(i) = N(i) - 1$.
\bigskip
\noindent
{\bf Theorem 17: Topology of the stability region, ${\bf R}_{\delta}$, for
${\bf F}_{t; \delta}$. (First ``Zipper Theorem'')} {\sl There exists a
strictly decreasing
sequence, $\delta_i$, with $\delta_i = 1-2^{-1/i}$, $i\geq 0$ such
that for $\delta \in (\delta_{i+1}, \delta_i)$, ${\bf R}_{\delta}$ has
$N(i) + N'(i)$ connected components, $N(i)$ of which are simply
connected, the others being infinitely connected ( {\it i.e.},
possessing infinitely many holes). Furthermore,
$A_{p/q}^{\oplus,stab}$ can intersect $A_{p'/q'}^{\oplus,stab}$, and
does intersect it for $\delta$ big enough, if and only if ${p\over q}$
and ${p'\over q'}$ are Farey neighbors.} \bigskip
\noindent
{\bf Proof of Theorem 17.}
The last statement follows trivially from the fact that the height of
$L_{p/q}^{+,stab}$ is a strictly decreasing function of $q$ (see the
proof of Theorem 12), and the definition of Farey neighbors. So let
${p\over q}$ be some rational number in $[0,1]$, and  ${p(i)\over q(i)}$
be the sequence
of its Farey neighbors on one side, with $q(i+1)>q(i)>q$: we want to
show that the value $\delta ({p\over q},{p(i)\over q(i)})$ of the
parameter $\delta$, at which $A_{p/q}^{\oplus,stab}$ and
$A_{p(i)/q(i)}^{\oplus,stab}$ detach from each other, decreases when
$i$ increases (hence a zipper effect, except for the case when $q=1$
where the detachment value $\delta=.5$ does not depend on $i$), with
the limit value $\delta_q$ a decreasing function of $q$ .

Let us consider the case when ${p(i)\over q(i)}>{p\over q}$, the other
case being treated in the same way. We consider a map $t_{\Omega ,S;
\delta}$ at the intersection of the right boundary of
$A_{p/q}^{\oplus,stab}$ with the left boundary of $A_{p(i)/q(i)}$, and
need to express the fact that $S$ corresponds to the top of the
stability region $A_{p(i)/q(i)}^{\oplus,stab}$. Let $T_{\omega ,S;
\delta}$ be the lift of $t_{\Omega ,S; \delta}$ such that the lifts of
${1+\delta\over 2}$ are fixed points under $T^q_{\omega ,S; \delta}$
(this choice can be made because we work on the right boundary of
$A_{p/q}^{\oplus,stab}$). Then the graph of $T^q_{\omega ,S; \delta}$
is a zigzag line creeping along the first diagonal; together with the
first diagonal, this graph determines a sequence of $q$ triangles per
interval of the form $I_n=[n+{1+\delta\over 2}, n+1+{1+\delta\over
2})$. The right triangle in $I_n$ has $x$-width $\delta +a_q$, and the
remaining ones are similar copies, with scaling factors ${1\over S}$
to  $({1\over S})^{q-1}$, hence
$$(\delta+a_q)\lbrack 1+{1\over S}+\dots +({1\over S})^{q-1} \rbrack
=1\,.$$ Each of these triangles has slopes $S^q$ and $sS^{q-1}$, and since
we are on the left boundary of the top of $A_{p(i)/q(i)}^{\oplus,stab}$, we
have $$\sigma S^{q(i)-1}=1\,.$$ Each triangle is followed (to the right) by
one which is either bigger or smaller, in the way the residue classes
$\bmod \,q$ are organized: the description of the organization of  residue
classes $\bmod \,q$ was performed by E.  B. Christoffel in the last century
[Ch], and it is explained in [STZ] how this relates to the kneading theory of
circle maps, as investigated (with a different name) one century before as
a question in number theory by Johan Bernoulli (the astronomer of the
famous family) [Be], and contemporarily by Marston Morse and Gustav
Hedlund [MH] in the context of symbolic dynamics. On the other hand,
because we are on the left boundary of $A_{p(i)/q(i)}$, the tips of the
triangle lift a periodic orbit with period $q(i)$: since ${p(i)/q(i)}$ and
${p\over q}$ are Farey neighbors, we get from the kneading theory of degree
two circle maps that there are either $n$ or $n+1$ points of the orbit of
${1-\delta\over 2}$ between two consecutive points of the orbit of
${1+\delta\over 2}$ ({\it i.e.}, in a triangle) [GoT]. More precisely, there
are
$n$ points if the triangle is smaller than the previous one, and $n+1$ points
otherwise. Putting together all this kneading information, basically easy
but too long to be recalled here in details (see [STZ], [GoT] and references
therein), the fact that the tip of one triangle
is the $n^{\rm th}$ or $(n+1)^{\rm th}$ image of the previous tip yields
$${a_q\over \delta}=\sigma S^{q(i)-1}\,.$$
Reassembling the equations gotten so far, we get $$\delta = {1\over
2}\cdot {S^{q-1}\over 1+S+\dots+S^{q-1}}\,.$$ This last expression
increases with $S$, hence the zipper effect, except for $q=1$. The limit
value $\delta_q$, corresponding to $S=S_c={1\over 1-\delta}$, gives an
explicit equation for $\delta_q$, to wit $$1/(2 \delta) = 1 + (1 - \delta) +
{(1 - \delta)}^2 + \cdots + {(1 - \delta)}^{q-1}\,,$$ hence $$\delta_q = 1-
2^{-1/q}\,.$$
\rightline {({\bf Q.E.D.}   Theorem 17.)}
\bigskip
The bifurcations at the first three $\delta _i$'s are illustrated by Figure
15.
\bigskip
A subset $A$ of the circle  is a {\it circularly ordered chaotic attractor}
for $f$, with rotation number $p/q$, if $A$ is a chaotic  attractor and for
some nicely ordered periodic orbit  $P_0, \dots, P_{q-1}$ with successive
indices around the circle, and with rotation number $p/q$,
$$x \in A  \cap I_i\,\Rightarrow \,f(x)  \in A \cap I_{ (i+p)_1 }\,,$$
where $I_i$  is the arc $(P_i, P_{ (i+1)_1 })$. An example of coexistence
of stable periodic orbit and circularly ordered chaotic attractor is
illustrated in Figure 16.

Let ${\bf R}^{\prime}_{\delta}$ be the set of   tip maps
$t_{\Omega ,S;\delta}$ that have a stable  periodic orbit or a
circularly ordered attractor.

\noindent {\bf Theorem 18: Topology of the
${\bf R}^{\prime}_{\delta}$'s. (Second ``Zipper Theorem'')}
{\sl
The statements of Theorem 17 apply to ${\bf
R}^{\prime}_{\delta}$ with a sequence $\delta_i^{\prime}$, where
$\delta_i^{\prime} > \delta_i$,  and with $A_{p/q}^{\oplus,stab}$
extended to include the maps in $A_{p/q}^\oplus$ with circularly
ordered chaotic attractors.   }
\bigskip
\noindent
{\bf Proof of Theorem 18.}
The proof goes essentially as for the preceding theorem, the only
difference being that the boundary of $A_{p(i)/q(i)}^{\oplus,stab}$
condition has to be replaced by a relation describing the boundary of
the circularly ordered chaotic region with rotation number
${p(i)\over q(i)}$. This equation is gotten by the geometry of the
graph of the restriction $t_{\Omega ,S; \delta}^{q(i)}$ to a $J_k$
(with the notations of the proof of Theorem 9), as depicted in Figure
17, and reads $${1\over S^{q(i)}}+{1\over \sigma S^{q(i)-1}}=1\,.$$
Combining this with the other relations yields  $$S^{q(i)+1}(1-
\delta)-S^{q(i)}-S+1=0\,,$$
hence
$$\delta = {1\over 2+{1\over S^{q(i)-1}}}\cdot {S^{q-1}\over
1+S+\dots +S^{q-1}}\,,$$ with the same limit value
$$\delta ''_q = 1-2^{-1/q}\,,$$
as before, but no control of the zipper effect yet, because of the
$q(i)$ dependence in the formula for $\delta$. However, since
$\delta$ decreases when $q$ increases, the following can be shown:

Let $R'_{p/q}$ be the closure of the subset of $A_{p/q}^\oplus$ where the
maps have stable periodic orbits or circularly ordered chaotic attractors.
As $\delta$ decreases to $\delta '_q$, the $R'_{r/s}$'s (where ${r\over
s}$ is
a Farey neighbor of ${p\over q}$) with bigger denominators detach
themselves from $R'_{p/q}$ before those with smaller denominators do. That
is, the unzipping takes place in the reverse order from that in Theorem 17.

\rightline{({\bf Q.E.D.}   Theorem 18.)}
The region ${\bf R}^{\prime}_{\delta}$ is shown in Figure 18 for $\delta =
\delta '_3$.
\bigskip
\centerline{{\bf 11. Summary, Discussion, and Open Issues}}

Since it is virtually impossible to summarize the results of eighteen
theorems and many observations in a concise manner in words, we will limit
our ``summary'' to the sketch in Figure 19, which illustrates many
of the phenomena we have described by displaying
the structure of parameter space for the tip
maps with $\delta >{1\over 2}$ around  $A_{1/2}$. We will turn in the
rest of this section to a discussion of the implications of our results
and of some important remaining open issues.

Apart from their interest as dynamical systems in their
own right, the piecewise linear circle maps have proven to be
interesting and valuable in several other contexts. First, as
a pedagogical example of a non-trivial, essentially solvable
model for mode locking and the quasi-periodic transition to
chaos, one can use them to clarify with explicit examples
and calculations concepts that require much deeper analysis
in the general smooth case. Second, our analytic
insight into piecewise linear maps has both stimulated conjectures
and suggested methods of proof that apply to the general
smooth case. Specifically, after a lecture given on this
material, a number of conjectures on the smooth case
were indeed proved [EKT] by Adam Epstein,
Linda Keen and one of the authors (C.T.), using a method initiated by
Bill Thurston in the context of interval maps. The [EKT] results
are parallel to (but slightly less strong) than those
we obtained in $\S$8 above in a mostly trivial manner for the plateau map
families. For the tip family, the counterpart to the results in [EKT] is
reasonably easy, but to go beyond, we needed a much deeper analysis
which was carried out by Marco Martens with two of us (R.G. and C.T.), as
reported in [GMT]. In any case, the piecewise linear maps have already
served well in this second role. Third, although in view of the seeming
non-generic
nature of the non-smooth map, one might expect that finding a physical
realization of phenomena such as the ``Arnold sausages'' would be
difficult; in fact, again following a lecture on this material,
preliminary numerical simulations at the University of Frankfurt [ZKM]
suggest that the phenomena should be experimentally
observable in driven Josephson junctions; further work on this
is in progress.

Among the other open issues, perhaps the chief matter left unresolved
by our study is the measure of the mode-locked intervals
below the critical line. Although our numerical data strongly suggest
that this measure is strictly less than one for all $\delta < 1$ and
for all $S < S_c$, we have at present no proof of this conjecture.

A second matter worth further study is motivated by adopting
the numerical analyst's perspective and considering
the piecewise linear maps as ``finite element, linear spline approximations''
to the smooth case. This view suggests the study of ``higher-order''
piecewise linear circle maps -- {\it i.e.}, multi-spline fits having
more than two independent slopes -- and, in principle, of studying the
convergence properties of the bifurcation sequences of these
maps to those of the smooth case as the number of splines increases.
Indeed, an analogous study [HC] comparing the (two-spline) tent and
(three-spline)
trapezoidal maps to the (four-spline) house map shows that, as expected,
the house map captures more of the bifurcation sequence of the logistic map
than its simpler cousins. However, the effort required to undertake detailed
studies of the multiple-parameter, higher-order piecewise linear maps
suggests that one should have a clear understanding of the importance of the
questions
being studied before proceeding. Further, a recent study [MT]
has shown rigorously that no finite segment PWL map with no zero slopes can
capture
the full renormalization sequence of the logistic map.

\bigskip
\centerline {{\bf Acknowledgements}}
\bigskip

Over the long course of this work we have benefited from interactions
with a large number of colleagues. A major mathematical motivation for
this project was the resistance to proof
of some conjectures about the standard family of circle
maps. As noted above, through discussions with  Adam Epstein, Linda Keen,
and Marco Martens, we not only gained insight into the present work
(as reflected in [GMT]) but were also able to prove some of these conjectures
in
[EKT]. It is thus a great pleasure to acknowledge the value
of our discussions with those colleagues, as well as
with Roy Adler, Kostya Khanin, Bruce Kitchens, Thomas Novicky and Michael Shub,
whose
patient listening helped us find and/or fix proofs for numerous statements
reported here. We also enjoyed many valuable discussions about the
phenomenon of mode locking in the physical world with Leon Glass,
Michael Mackey, Ronnie Maineri, and Steve Strogatz. We are grateful
to Werner Martienssen, Uwe Kr\"uger, and Holger Zimmerman for sharing
with us their preliminary observation of ``Arnold sausages'' in a
model of a driven Josephson junction.

\filbreak
\bigskip
\centerline {\bf {Appendix A: Essential Definitions and Background Material.}}
\bigskip

This Appendix assembles some background material on the dynamics of
circle maps, interval maps, and topological entropy: its only use for experts
is to establish the notation we use in the rest of the paper. Non-experts
are advised to sketch figures, wherever appropriate, of each of the
definitions,
concepts, and arguments, as this will greatly improve their understanding
of the paper. As in the main
text, all fractions will be written in lowest order terms, except where
otherwise
specified, and we shall use the notation $(z)_n\buildrel \rm def\over =
z\,\bmod\,n\,\,.$
\bigskip
\noindent
{\bf {A-1. Circle maps, lifts, and degree.}}
\bigskip
When dealing with the dynamics of circle mappings, it is useful to consider the
circle as
the set of real numbers $\bmod\,1$, {\it i.e.}, to think of the circle as being
the
set $\TT=\RR /\ZZ$, meaning that we consider two real numbers $x$,
$y$ in the set of real numbers $\RR$ to be {\it equivalent} if and
only if they differ by some integer $n$ in the set $\ZZ$ of rational
integers. This means that our angles are measured in the unit $2\pi\,
radians$.

The circle $\TT$ is then the set of the
equivalence classes so defined, a good set of representatives of these
equivalence classes being the semi-open interval $[\,0,\,1)$.  The
topology of the circle is obtained by endowing $\TT$ with the
distance $d$ defined by: $$d((x)_1\,,\,(y)_1)=\min(|(x)_1-
(y)_1|,\,|(x)_1| +|1-(y)_1|,\,|(y)_1|+|1-(x)_1|)\,.$$

To each continuous circle map $f:\TT\to \TT$, we can associate {\it
lift
maps}, where the continuous map $F:\RR\to \RR$ is a lift of $f$ if and
only if $$(F(x))_1=f((x)_1)\,\,.$$
There is a countable infinity of such lifts for any $f$: if $F_0$ is
one of the lifts  of $f$, the set of lifts of $f$ is the set of maps
$F_n=F_0+n$, where $n\in\ZZ$. On some occasions, and in
particular for the purpose of drawing pictures, it is easier to
suppose that $F_0(0)$ belongs to the interval $[\,0,\,1)$.
However, this convention cannot
hold when we interpolate the successive lifts $F_n$ of some maps
$f$, by a continuous one parameter family $\lbrace F_\mu\rbrace$
with $F_{(\mu+1)}=F_\mu+1$.

If $F$ is the lift of a continuous circle map $f$, there exists an
integer $d$ such that $$F(x+1)=F(x)+d\,\,,$$ for all real numbers $x$.
This number $d$ is called the {\it degree} of $f$ (or of $F$), and a
good intuition of its meaning can be gotten by contemplating the
graphs for maps with degrees respectively equal to $-1$, $0$, $1$ and
$2$ in Figure A1: $d$ is the algebraic number of times the circle is
wrapped onto itself by the map.

The rotations $R_\theta(x)=(x+\theta)_1$ have degree one. Since the
degree varies continuously for continuous deformations of the maps,
and since we are interested in parametrized continuous families
containing the rotations, we only consider degree one in the paper and
the rest of the Appendix.  Hence, throughout the main text and the
rest of the appendices {\it circle map} will always mean degree-one
continuous circle map, and a
map will be called a {\it lift} if and only if it is the lift of a degree one
circle map.
\bigskip
\noindent
{\bf {A-2. Rotation intervals and rotation numbers.}}
\bigskip
Let $f$ be a circle map, and $F$ be a lift of $f$.
Using the notation $F^n(x)$ for the $n^{th}$ iterate
of $F$, we set
$$\underline{\rho}_F(x)=\liminf_{n\to\infty}{F^n(x) \over
n}\,\,,$$ $$\overline{\rho}_F(x)=\limsup_{n\to\infty}{F^n(x)
\over n}\,\,,$$ and define the {\it rotation interval} of $F$
[NPT] as $$I(F)=[\,a,\,b]\,\,,$$
where
$$a=\inf_{x\in
\RR}\underline{\rho}_F(x)\,\,,\,\,b=\sup_{x\in
\RR}\overline{\rho}_F(x)\,\,.$$

In some cases, and in particular when $F$ is non-decreasing,
$I(F)$ reduces to a single number, also denoted  $\rho (F)$, called the
{\it rotation number} of $F$ [Po]; then the $\limsup$ and $\liminf$
can be replaced by a limit in the defining formulas. In this case,
$(\rho (F))_1$ is called the {\it rotation number of $f$}.
Similar passing from $F$ to the circle map $f$ itself
is in general impractical because the
rotation interval of a general lift may be longer than one. Intuitively,
the uniqueness of the rotation number when $F$ is non-decreasing can
be pictured by what happens when cars race on a circuit where the
road is too narrow to allow passing.

When $I(F)$ is a single point, $f$ has a periodic orbit if and only if
$\rho (F)$ is a rational number.  Furthermore, if $F$ is non-decreasing
with $\rho (F)= {p\over q}$, the period of all periodic
orbits of $f$ is equal to $q$. This number $q$ is not alone sufficient to allow
a
combinatorial description of the dynamics of $f$ restricted to these
periodic orbits: with ${P\over Q}=({p\over q})_1$, $P-1$ is the number of
points of the orbit are jumped over from a point to its image, in the
positive direction around the circle (equivalently, $P$ is the number of
revolutions around the circle which is needed to come back to any point
when following the orbit). Thus this number is constant on the points
belonging to all periodic orbits of $f$ when $F$ is non-decreasing with
rational rotation number ${p\over q}$. In other words, the dynamics
of such a map $f$, restricted to any of its periodic orbits, is
combinatorially the same as the dynamics of the rotation by ${P\over Q}$. The
number $p$ is also the number of orbits of $F$ (forward and backward)
which project to any given periodic orbit of $f$ by $\bmod\,1$.

Non-decreasing maps form an important class in the study of degree-one
endomorphisms. Following Alain Chenciner, we designate as {\it
fomeomorphisms} (a French contraction for ``false homeomorphism'') the
continuous degree-one circle maps whose lifts are non-decreasing. A
periodic orbit of a circle map, which is also a periodic orbit of some
fomeomorphism with rotation number ${P\over Q}$ (hence
combinatorially like an orbit of the rotation by the angle ${P\over Q}$)
is called {\it ${P\over Q}$-ordered}. More generally, an orbit of a
circle map, which is also an orbit of some fomeomorphism with
rotation number $\Omega$, is called {\it $\Omega$-ordered}. Lifts of
such orbits are then said to be {\it $\omega$-ordered} for some
$\omega$ such that $\Omega= (\omega)_1$. In particular, a ${p\over
q}$-ordered orbit is always the lift of some periodic orbit, invariant
under a fomeomorphism with rotation number ${P\over Q}=({p\over
q})_1$. We sometimes say {\it nicely ordered} for ``$\omega$-ordered
for some $\omega$".

We denote by ${\cal F}^0(\RR)$ the space of lifts of fomeomorphisms
equipped with the {\it sup norm}: {\it i.e.}, two such maps are close
together if their respective values at each point of an interval of
length one are close to each other. More generally, ${\cal F}^k(\RR)$
stands for the space of $k$ times continuously differentiable lifts of
fomeomorphisms, where two of them are close together if the maps
and their $k$ first derivatives are.
Further, we shall use the following
definitions for inequalities between real maps: $$F>G \Leftrightarrow
\forall x,\,F(x)>G(x)\,,$$
and
$$F\geq G \Leftrightarrow \forall x,\,F(x)\geq G(x)\,.$$
With these definitions and preliminaries, we next state some
classical results about the rotation number of non-
decreasing maps, which go back to Henri Poincar{\'e} [Po] in the
context of degree-one homeomorphisms and are not harder to prove
in our slightly generalized context.
\bigskip
\noindent
{\bf Theorem A.} {\sl

\noindent
- (i). For $F$ and $G$ in ${\cal F}^0(\RR)$,
$$F\geq G\qquad\Rightarrow\qquad \rho (F)\geq \rho(G)\,\,,$$
$$F>G\,\,\&\,\,\rho(F)\,\,or\,\,\rho(G)\,\,irrational\quad\Rightarrow
\quad \rho (F)> \rho(G)\,\,.$$

\noindent
- (ii). The rotation number, as a function $\rho:{\cal F}^0(\RR)\to
\RR$ is continuous.

\noindent
- (iii). For $F$ and $G$ in ${\cal F}^0(\RR)$, if $\rho(F)$ is irrational
and $f$ has a dense orbit, then, $$F\geq G\quad and\quad
F\not= G\qquad\Rightarrow\qquad \rho (F)>\rho(G)\,\,.$$ }
\bigskip
\noindent
{\bf Corollary A'.} {\sl For a family $\lbrace F_{\mu}\rbrace$ in
${\cal
F}^0(\RR)$, defined by $F_{\mu}=F_0+\mu$, the rotation number of
$F_{\mu}$ is a non-decreasing function of $\mu$.}
\bigskip
If $F$ is a degree-one lift, we denote by $F^+$ its monotone upper-
bound, and by $F^-$ its monotone lower-bound (see Figure A2). In
formulas: $$F^+(x)=\sup_{y\leq x}(F(y))\,,$$
$$F^-(x)=\inf_{y\geq x}(F(y))\,.$$
\bigskip
\noindent
{\bf Theorem B [CGT, Mi2].} {\sl For any lift $F$,

\noindent
- (i). $I(F)=[\,\rho(F^-),\,\rho(F^+)]\,.$

\noindent
- (ii). Furthermore, for each $\omega\in I(F)$, there is a non-
decreasing lift $F_\omega$ which coincides with $F$ where it is not
locally constant, and with  $\rho(F_\omega)=\omega$.}
\bigskip
\noindent
{\bf Corollary B'.} {\sl  For any lift $F$, if  $\omega\in I(F)$, then
F has a $\omega$-ordered orbit.}
\bigskip
\noindent
{\bf {A-3. Denjoy Theory.}}
\bigskip
The following result about smooth maps will be used in some proofs:
it is a weakened form of the ``famous'' result by Arnaud  Denjoy [De]:
\bigskip
\noindent
{\bf Theorem of Denjoy.} {\sl For $F$ in ${\cal F}^2(\RR)$,
if $\rho(F)$ is irrational then $f$ has a dense orbit. }
\bigskip
\noindent
{\bf Remark.} Bohl [Boh], and independently Denjoy [De],
constructed counterexamples to this statement in ${\cal
F}^1(\RR)$. Such maps usually called {\it Denjoy
counterexamples}.
\bigskip
\noindent
{\bf {A-4. Topological entropy.}}
\bigskip
Let $M$ be a compact metric space with distance $d$, $f$ an {\it
endomorphism} of $M$ ({\it i.e.}, a continuous map from $M$ to itself).
For a positive real number $\epsilon$ and an integer $n$, the subset $S$
of $M$ is {\it $n$-$\epsilon$-separated} if for each pair $(x,\,y)$ of
distinct points of $S$, there is an $m$ in $\lbrace 0,\,1,\dots ,\,n-
1\rbrace$ such that $$d(f^m(x),\,f^m(y))>\epsilon\,.$$ Let
$N(n,\,\epsilon)$ stand for the maximal cardinal ({\it i.e.}, the number
of elements) of an $n$-$\epsilon$-
separated set. We set $$H(\epsilon)=\limsup_{n\to \infty}{1\over
n}\,log\,N(n,\,\epsilon)\,.$$ Then the {\it topological entropy} ([AKM],
[Bow]) of $f$ is $$h(f)=\lim_{\epsilon\to 0}H(\epsilon)\,.$$
The topological entropy, originally devised as an invariant of
topological conjugacy [AKM], is also a measure of the dynamical
complexity of a map. By a Theorem of [MSz], for a piecewise monotone
endomorphism $f$ of the circle or the interval, the topological entropy
$h(f)$ is given by: $$h(f)=\limsup_{n\to\infty } {1\over n} log(N(n))\,,$$
where $N(n)$ stands
for the number of maximal arcs or intervals where $f^{\circ n}$ is
monotone,
or the number of isolated periodic points with period $n$ of $f$.
Furthermore, we have the following results:
\bigskip
\noindent
{\bf Theorem C [BFr, Mi1].} {\sl
\noindent
An endomorphism of the interval has positive topological entropy if
and only if it has a periodic orbit whose period is not a power of two.
} \bigskip
\noindent
{\bf Theorem D [Mi1].} {\sl
\noindent
An endomorphism $f$ of the interval I or the circle $\TT$ has
positive topological entropy if and only if it has a horseshoe, {\it
i.e.}, for some $n$ and some interval $J\subset I$, or some arc
$J\subset \TT$, $J$ contains two disjoints subintervals $J_0$ and
$J_1$ such that both $f^n(J_0)$ and $f^n(J_1)$ contain $J$. }
\filbreak
\centerline
{\bf {Appendix B: Stunted families.}}
\bigskip

Figures A3 and A4 explain respectively how to construct {\sl stunted
families} out of maps on the interval and the circle. Such stunted
families are particularly easy to study since the topological entropy,
and finer topological invariants as furnished by kneading theory [MTh],
vary monotonically and continuously in each parameter ({\it cf}
[DGMT]). Here we describe some examples on the interval which will
help us in the study of some special circle maps.

For $S>1$ and $\mu \in [0,1+{1\over S}]$, let $f_{S,\mu}$ be the
endomorphism of the unit interval defined by
$$ f_{S,\mu}(x)\,\,=\quad\left\{ \eqalign {\min (Sx, 1)
\quad if\,\,\,\,0 \leq x \leq {\mu\over 2}\,,\cr
\max(\min (S(\mu-x),1),0)\quad if\,\,\,\,{\mu\over 2}
\leq x \leq
 {S(\mu+1)-1\over 2S}\,,\cr \max (S(x-1)+1,0) \quad
if\,\,\,\,
 {S(\mu+1)-1\over 2S} \leq x
	\leq 1\,,\cr
}\right .$$
This map fixes the end points, and has two extremal values (hence is
{\sl bimodal}), a maximum  $$M_{S,\mu}=\min (1,{S\mu\over 2})\,,$$
and a minimum
$$m_{S,\mu}=\max (0,S{\mu-1\over 2}+{1\over 2})\,,$$
so that the graph is like a hill followed by a valley, except in the
limit cases $\mu\in \lbrace 0,1+{1\over S} \rbrace$ where the hill or
valley degenerates to a point. We call {\sl plateaus} the segments
where the extrema are attained: in some cases, one or both plateaus
may be reduced to a single point.

Fixing $S>1$ and $\mu \in [0,1]$, for $a\in
[0,M_{S,\mu}]$ and $b\in [0,1-m_{S,\mu}]$ such that $a+b\geq 1$, we
have a two-parameter stunted family defined by
$$ c_{S,\mu;a,b}(x)\,\,=\quad\left\{ \eqalign {\min(f_{S,\mu}(x),a)
\quad {\rm under\ the\ hill}\,,\cr \max(f_{S,\mu}(x),1-b)
\quad {\rm in\ the\ valley}\,,\cr
}\right .$$
The $(a,b)$-parameter space of $c_{S,\mu;a,b}$ is represented in Figure
A5. A nice feature of these stunted families is that, since $S>1$, one has the
\bigskip
\noindent
{\bf Density Property.} {\sl The unions of all
preimages of the plateaus ({\it i.e.}, under all iterates of the map) are
dense in $[0,1]$.}
\bigskip
\noindent
A proof of the density is easily supplied. A little more work is needed
for the following result (see, e.g., the appendix of [BMT]).
\bigskip
\noindent
{\bf Measure Property.} {\sl If any of the plateaus has non-zero
length, the unions of all preimages of the plateaus ({\it i.e.}, under all
iterates of the map) have Lebesgue measure 1.}
\bigskip
The combinatorial structure of an orbit is determined by the sequence
of segments of monotonicity of the map visited by the orbit, as
formalized in the kneading theory of Milnor and Thurston [MTh]. In
order to limit the length of this paper, we shall avoid here explicit
use of kneading theory (and its language), but a good knowledge of it
would certainly simplify the comprehension of the rest of this
section and its applications: for the kneading theory of interval maps,
the reader is refereed to [MTh], and for circle maps to [AM1]. We will
say the combinatorial behavior of the orbit of a plateau is {\sl
plateau-avoiding} if the orbit of the plateau does not hit any plateau.
The following property is easy to check.
\bigskip
\noindent
{\bf Parallel Property.} {\sl Any plateau-avoiding combinatorial
behavior $\beta$ of the orbit of a plateau occurs on a subset
$s_\beta$ of the $(a,b)$ parameter space,  which is a segment
parallel to the $b$-axis, with one end point on the $a$-axis, for the
left plateau, and a segment parallel to the $a$-axis, with one end
point on the $b$-axis, for the right plateau.}
\bigskip
We just mention here the following:
\bigskip
\noindent
{\bf Corollary 1.} {\sl For a dense subset $D$ of the $(a,b)$ parameter
space, one point of a plateau belongs to a periodic orbit.}
\bigskip
\noindent
{\bf Remark.} With the methods in [BMT], one can show that in $D$, the
orbits of almost all points in $[0,1]$ converge to a periodic orbit
which has (at least) one point in a plateau, and that the complement
of $D$ in the $(a,b)$ parameter space, has zero Lebesgue measure.
More important for our purpose is the next consequence of the
Parallel Property.
\bigskip
\noindent
{\bf Isolated Points Property.} {\sl Any pair of plateau-avoiding
combinatorial behavior $(\beta, \beta ')$ of the orbits of the plateaus
occurs for at most a single point $P_{(\beta, \beta ')}$ of the $(a,b)$
parameter space.}

Combining the Parallel Property and the Isolated Points Property with
the combinatorial theory for the boundary of chaos for bimodal maps
([MaT2], [MaT3], [Mu]), we get the following
\bigskip
\noindent
{\bf Theorem E .} {\sl The region corresponding to zero topological
entropy in the $(a,b)$ parameter space of $c_{S,\mu;a,b}$, is a closed
set, separated from the rest of the parameter space by a curve of
length bounded from above by $2$, made of countably many pieces of
straight lines crossing the gaps of a Cantor set embedded in the two-
dimensional parameter space.}
\bigskip
A construction of the curve discussed in Theorem E is shown schematically in
Figure A6. Now, fixing $S$, we can form three-parameter families
(with parameters $\mu$, $a$ and $b$), such that the parameter space
is as represented in Figure A7-$\alpha$ when $S>3$, and as in Figure A7-$\beta$
otherwise: we denote by ${\bf C}_S$ the 3-cell which parametrizes
the family $c_{S,\mu;a,b}$ for given $S>1$. Some two-dimensional
surfaces in these cells will occur in the study of some families of
circle maps. To get a result similar to Theorem E in these surfaces,
we will use the following monotonicity result .
\bigskip
\noindent
{\bf Theorem F.} {\sl If some plateau-avoiding behavior of the right
plateau occurs both for $c_{S,\mu;M_{S,\mu},b}$ and $c_{S,\mu
';M_{S,\mu'},b'}$ for $\mu'>\mu$, then $b'< b$. Similarly, if some
plateau-avoiding behavior of the left plateau occurs both for
$c_{S,\mu;a,m_{S,\mu}}$ and $c_{S,\mu ';a',m_{S,\mu'}}$ for
$\mu'<\mu$, then $a'< a$.}
\bigskip
\noindent
{\bf Proof of Theorem F.} The second statement is equivalent to the
first one, that we deduce from the Measure Property as follows.

\noindent
- From kneading theory, we know that  $c_{S,\mu;M_{S,\mu},b}$ and
$c_{S,\mu ';M_{S,\mu '},b'}$ have the same set of preimages of the
plateaus (as labeled by the successive branches of the map where the
preimages are taken). This is sufficient to conclude when
$M_{S,\mu}<1$.

\noindent
- When $M_{S,\mu}=1$, we conclude thanks to the following Lemma.
\bigskip
\noindent
{\bf Lemma F'.} {\sl For any map of the form $c_{S,\mu;1,b}$, with
$b<1$, and all $n\geq 0$, there is an injective map from the set of
intervals which are $n^{\rm th}$ preimages of the left plateau to the
set of intervals which are $n^{\rm th}$ preimages of the right
plateau, and the injection is strict for some values of $n$.}

\rightline {({\bf Q.E.D.} Theorem F.)}
\bigskip
{\bf Proof of Lemma F'.} From the already mentioned property in
kneading theory, we choose to work with the map $c_{3,{2\over
3};1,1}$ for convenience of the exposition, without loss of generality.
Under this map, the two turning points, ${1\over 3}$ and ${2\over 3}$,
have symmetric sets of preimages, and we only need to show that for
any  map $c_{3,{2\over 3};1,b}$  with $b<1$, the left turning point has
lost at least as many preimages as the right turning point, at each
generation, and sometimes more.

We will only consider preimages of ${1\over 3}$ and ${2\over 3}$ to
the left of ${2\over 3}$, preimages on the other side of ${2\over 3}$
being treated similarly. For each $n>0$, let $x_{n,N(n)}<x_{n,N(n)-
1}<\dots <x_{n,1}$, with $x_{n,1}<{2\over 3}$, be the preimages of
${2\over 3}$, up to the $n^{\rm th}$ generation, and such that
$c_{3,{2\over 3};1,1}(x_{n,i})<1-b$, and let $m_{n,i}\leq n$ be the
generation index of
$x_{n,i}\,$, {\it i.e.}, $c_{3,{2\over 3};1,1}^{m_{n,i}}(x_{n,i})={2\over 3}$.
Let $I_{n,1}$ stand for the interval between $x_{n,1}$ and ${2\over 3}$.
Then $c_{3,{2\over 3};1,1}^{m_{n,1}}(I_{n,1})$ contains the other turning
point ${1\over 3}$, so that $I_{n,1}$ contains a preimage of ${1\over 3}$
with generation index at most $m_{n,1}\leq n$. Similarly, for each $i\in
\lbrace 2,3,\dots,N(n)\rbrace$, the interval $I_{n,i}=(x_{n,i},x_{n,{i-1}})$
contains a preimage of  ${1\over 3}$ with generation index at most
$\max(m_{n,i},m_{n,{i-1}})\leq n$, as we show now to conclude the proof
of injectivity.

- either $m_{n,i}=m_{n,{i-1}}$, in which case the paths of the orbits of
$x_{n,i}$ and $x_{n,{i-1}}$ until ${2\over 3}$ is reached have to be
different: the only way this can occur is the image of $I_{n,i}$, under
$c_{3,{2\over 3};1,1}^q$ for some $q<m_{n,i}$, contains ${1\over 3}$
since $x_{n,i}$ and $x_{n,{i-1}}$ are consecutive.

- or  $m_{n,i}\neq m_{n,{i-1}}$, that we rewrite as $m_{n,j}<m_{n,k}$:
then, when the orbit of $x_{n,k}$ reaches ${2\over 3}$, the orbit of
$x_{n,j}$ has reached  the fixed point $0$, and again, some image of
$I_{n,i}$ contains ${1\over 3}$.

To show that the injection is strict for some values of $n$, it remains
to show that for some values of $n$, there is a preimage of ${1\over
3}$ with generation at most $n$ to the left of $x_{n,N(n)}$. Notice then
that either

- (i) $x_{n+1,N(n+1)-1}=x_{n,N(n)}$, or

- (ii) $m_{n+1,N(n+1)-1}=n+1$.

In case (ii), we conclude as before that the image of $I_{n+1,N(n+1)}$,
under $c_{3,{2\over 3};1,1}^q$ for some $q<m_{n+1,N(n+1)}$, hence
$q\leq n$, contains ${1\over 3}$.

In case (i), if $x_{n+1,N(n+1)}$ is not the preimage of $x_{n,N(n)}$, we
conclude as in case (ii), and $x_{n+1,N(n+1)}$ cannot be the preimage
of $x_{n,N(n)}$  for all $n$'s, since such monotonic sequences would
have to accumulate on one of the end points of the interval, and the
leftmost preimage of order $n$ of ${1\over 3}$ is to the left of the
leftmost preimage of ${2\over 3}$ or order $n$ or less.
\rightline {({\bf Q.E.D.} Lemma F'.)}

\noindent
{\bf Remarks.}

- Theorem F has just been stated here with the generality required for
the study of the boundary of chaos: to suppress the words ``plateau-
avoiding" in this statement, one can use the methods in [BMT]. The
plateau-avoiding case is simpler since all preimages of both plateaus
are pairwise disjoint.

- If some plateau-avoiding behavior $\beta$ of the right plateau
occurs for $c_{S,\mu_\beta (b);1,b}$ (respectively if some plateau-
avoiding behavior $\beta$ of the left plateau occurs for
$c_{S,\mu_\beta (a);a,1}$) it follows from the proofs of Lemma F' and Theorem F
that $\mu_\beta (b)$ (respectively $\mu_\beta (a)$)
is a linear function.
Combining this with the previous remark, the same
linear property holds true if $\beta$ corresponds to a
periodic behavior of an end-point of a plateau.

\filbreak

\bigskip \noindent
{\bf Appendix C:  Computation of Boundaries of Some Regions
in Parameter Space.}

\bigskip
Since the maps discussed in this paper are piecewise linear,
it is possible to find exact equations for the boundaries of
the various parameter plane regions discussed
in the previous parts of the
paper, although in most cases obtaining such equations in a
simplified form is not trivial.  In this appendix, we will
describe the exact equations of most of the boundaries,
especially for the tip family, and give a brief
indication of how the equations were obtained.  The derivations
of the boundary equations are elementary and come from
a few simple facts about circle maps $f$ and their lifts $F$,
such as the following where we have identified the circle
with the interval $[0,1)$, $x \in [0,1)$, and $F$ is the
particular lift of $f$ satisfying $F(0) \in [0,1)$:

\medskip
   {\it -i)}  $x$ belongs to a $p/q$-cycle of the
circle map $f$ if and only if $F^q(x) = x+p$.

\medskip
   {\it -ii)}  On the left (respectively, right) boundary
of $A_{p/q}$ the graph of the
$q^{th}$ iterate $F^q$ of the appropriately chosen
lift is tangent to the line
$y=x+p$ from below (above), in the $(\Omega,U)$ parameter plane.

\medskip
   {\it -iii)}  On the left (respectively, right)
boundary of $A_{p/q}$, a $p/q$-cycle of a tip or
plateau map must
contain the first (respectively, last) turning point
of the map.  These turning points
are the first and last points in Figure 1 where the slope
is undefined.

\medskip
   {\it -iv)}  If $x$ is a member of a stable (attractive)
$p/q$-cycle of $f$ then at $x$ the slope of $F^q$ is less
than 1 in magnitude.

\bigskip \noindent 
{\bf C-1.  Boundaries of $A_{p/q}$ for the Tip Maps.}

\bigskip
In the sections below we will obtain the boundary
equations of each connected component of the interior
of the sausage-like structure $A_{p/q}$.  The points that
separate these connected components are called {\it nodes}.
The equations representing the boundaries of $A_{p/q}$
will change from one component to another of
$ {\it int}( A_{p/q} )$.  For each $p/q$ the components of
$ {\it int}( A_{p/q} )$ belong to disjoint
horizontal strips in the $(\Omega,S)$ plane.  Since the number
of components is finite, exactly one of them is unbounded.
This unbounded component, which is the component that intersects
the critical line, we call the {\it top} component of
$ {\it int}( A_{p/q} )$.
The {\it top node} is the node that
separates the top component from the rest of $A_{p/q}$.

\bigskip           
   {\bf C-1.1.  Nodes.}

\medskip
   In this section we will find the heights of the nodes in
the $(\Omega,S)$ parameter plane.
As discussed in the proof of Theorem 4 in  $\S$7, the
tongue $A_{p/q}$ has zero width in the $(\Omega,S)$ parameter
plane, for fixed $\delta$, when the corresponding circle map
$f$ satisfies $f^q = {\rm Id}$ (the identity map
on the circle).  Such a point $(\Omega,S)$ in the
parameter plane is a node, except at the
foot of the tongue where $S=1$.
Thus the nodes are the points that separate
the different ``links'' of the sausage-like structure $A_{p/q}$.

By setting the slope of $f^q$ equal to 1, as in the proof of
Theorem 4, one comes up with the polynomial equation that
must be satisfied by the parameter $S$ at a node:
$$ S^{q-k}[1 - S(1-\delta)]^k - \delta^k = 0,
                                     {\it ~~~~~~~~~~~(C1)} $$
for $k = 1, \dots, \lceil \delta q \rceil - 1$, where
$\lceil x \rceil = {\it ceiling}(x)$ = the least integer
greater than or equal to $x$.

At a node of $A_{p/q}$ the $p/q$-cycle for $t_{\Omega,S;\delta}$
must contain both turning points of the map.
As an illustration of how the nodes occur, consider Figure C1,
which shows how the members of a typical 2/5-cycle vary as
$(\Omega,S)$ descends along the left boundary of $A_{p/q}$
from the critical line $S=S_c$ to the foot of the tongue at $S=1$.
In this figure, $\delta \approx 1/2$, the circle is represented
as the interval [-1/2,1/2), and the portion of the circle
between the turning points $t_1 = -t_0 = (1-\delta)/2$ is the
big slope region of the map.  When a member of the 2/5-cycle
coincides with $t_0$ a node occurs.  This figure illustrates
the proof of Theorem 4 and also illustrates the reason that
the number of nodes in $A_{p/q}$ is $\lceil \delta q \rceil - 1$.

Equation (C1) can be used to determine the nodes of $A_{p/q}$
to any desired degree of accuracy with the help of Newton's
Method, and for a given $S$ by testing the left side of
(C1) for $k=0,1,\dots$, one can determine which link of the
sausage is cut by the horizontal line at $S$. Figure C2
illustrates typical locations of the turning points $t_0, t_1$ as
well as the big slope region of a tip map.
\vfill
\eject
\bigskip
{\bf C-1.2.  Further properties of the nodes.} 

\medskip
A summary of other properties of the set of nodes in
$A_{p/q}$ is given below.  These properties can be proved
by elementary methods and are illustrated by the various
figures in this section (see especially Figures C3 to C15,
which will be discussed in detail later).

\medskip \noindent
{\it Nodes in relation to the critical line:} 

\noindent
All nodes of $A_{p/q}$ are below the critical line.

\medskip \noindent
{\it Nodes for even $q$:} 

\noindent
If $\delta > 1/2$ and
$S_0 = \delta / (1-\delta)$ then all $A_{p/q}$ with even
$q$ have nodes on the line $S=S_0$.  In fact, the $p/q$
sausage has its ${q \over 2}{\it th}$ node on this line,
counting down from the top node.  If $\delta \le 1/2$ there is
no such $S_0$.

\medskip \noindent
{\it Rows of nodes:} 

\noindent
Assume $q>1$ and let
$t_1 = -t_0$  $ = (1-\delta)/2$.  Below the critical line
in the $(\Omega,S)$ plane,
the line $\Omega + S t_1 = t_0 + 1$ intersects $A_{p/q}$
at nodes only.  Similarly, there are rational curves
$\Omega = P(S)/Q(S)$ of all degrees ($k$ over $k-1$) that
intersect the sausages below the critical line only at nodes.

These rational curves are obtained by
solving $T^q_{\Omega,S;\delta}(t_1) = t_0 + p$
for $\Omega$ where $q = 1,2, \dots$ and where $T$ is the
lift of $t$.
Two such rational curves through nodes
are drawn in Figure C16 for $q = 1, 2$ and they can also
be seen in some of our other figures showing groups of
$p/q$-tongues.

\medskip \noindent
{\it How the nodes vary with $\delta$:} 

\noindent
Suppose $p/q \in [0,1)$ is in lowest terms.  Then:

  {\it a)~~}
For small $\delta$, $R_{p/q}$ has no nodes.

  {\it b)~~}
For $\delta$ sufficiently close to 1, $R_{p/q}$
has $\lceil q \rceil - 1$ nodes, i.e., it has its full
allowance of nodes and will get no more.

  {\it c)~~}
As $\delta \rightarrow 1$, all nodes of
$R_{p/q}$ approach the critical line.  (Recall that the
height $S_c$ becomes unbounded as $\delta \rightarrow 1$.
The upper nodes (those
with node number $k<q/2$) approach the critical line
absolutely:  $S_{\rm node} - S_c$ $\rightarrow 0$ as
$\delta \rightarrow 1$ while the other nodes approach
the critical line relatively:  $(S_{\rm node} - S_c)/S_c$
$\rightarrow 0$ as $\delta \rightarrow 1$.  As
$\delta \rightarrow 1$, the $\Omega$-coordinates of all
nodes approach 1/2, as illustrated in Figures C3 to C15.

\bigskip        
{\bf C-1.3.  Boundaries of $A_{p/q}$ above the top node.}

\medskip
   We now proceed to find equations of the exact boundaries
of $A_{p/q}$ for the tip family.  For simplicity we first
discuss these boundaries above the top node.  The top
component of ${\it int}(A_{p/q})$ intersects the critical
line, so that the boundary equations obtained in this
section will also yield the exact endpoints of the stability
intervals on the critical line.

Below the critical line the equations of the left and right
boundaries of $A_{p/q}$ will change from link to link of
the sausage structure.  In order more easily to understand
how the boundary equations are obtained we consider in Figure C2
a typical lift of the tip family.  The particular
lift shown corresponds to a parameter pair $(\Omega,S)$
above the critical line, where the small slope $s =
[1-S(1-\delta)]/ \delta$ is negative.  Above the first
node, the tip family boundary equations for $A_{p/q}$ will
be the same whether $(\Omega,S)$ is below or above the
critical line.

The three linear functions $f_1, f_2, f_3$
that we will use to define a lift of the tip map
$t_{\Omega,S;\delta}$, as shown in Figure C2, are:

\medskip
 $ f_1(x)  =  \Omega + s(x+1/2) - 1/2$,

 $ f_2(x)  =  \Omega + Sx$,  {\it \hskip2in (C2)}

 $ f_3(x)  =  \Omega + s(x-1/2) + 1/2,  $

\medskip \noindent
where for convenience in the following discussion we
will now identify the circle with the interval $[-1/2,1/2)$.
(We could use any interval of length 1; this one seems
most convenient.)
To help understand the three functions $f_1, f_2, f_3$,
recall that the map $t_{\Omega,S;\delta}$, for fixed
$\delta$, consists of two linear parts, one with the
``big slope'' $S$ and the other with the ``small slope'' $s$.
The function $f_2$ represents the ``big slope'' part and
the other two functions $f_1$ and $f_3$ represent the ``small
slope'' part of $t_{\Omega,S;\delta}$.  No matter where
$(\Omega,S)$ lies in the sausage structure $A_{p/q}$, we
will be able to represent $t_{\Omega,S;\delta}$ and all
its iterates using the three functions $f_1, f_2, f_3$ along
with the decrement function described below.  Note that $f_3$
is just a shifted version of $f_1$ since $f_3(x) \equiv $
$f_1(x-1) + 1$.

The above three functions will now be used to find the boundary
equations of $A_{p/q}$.  We first consider the left
boundary of $A_{p/q}$ above the top node.  On the left
boundary, the following equation must be satisfied:
$$ t_{\Omega,S;\delta}^q(t_1) = t_1, {\it ~~~~~~~~~~~~~~~(C3)} $$
where appropriate choices among $f_1$, $f_2$, $f_3$ must
be made for each of the iterates in $t^q$.
The thing that distinguishes one link of the sausage
from the other links
(see Figure C1) is the number of $p/q$-cycle points that
are outside the big slope region of the circle map.
To be more precise, if
the circle is represented by the interval $[-1/2, 1/2)$
then the ``big slope'' region is the interval $(t_0,t_1]$ in
Figure C2.  When no points of the $p/q$-cycle are outside this
interval then the parameter pair $(\Omega,S)$ is in the top
link, which includes the part of $A_{p/q}$ above the
critical line.
When k points of the cycle are outside the interval
then $(\Omega,S)$ is in the $k^{th}$ link from the top, where
$k = 1, \dots, \lceil q\delta \rceil - 1$ as shown by
Theorem 4.

In order to simplify the boundary equation, we introduce the
{\it decrement} function:
$$ d(x) = x-1. $$
Since the values of the lift wrap around the circle $p$ times
during a $p/q$-cycle, the left side of equation (C3) is
represented by $q$ copies of $f_2$ interspersed with $p$ copies
of the decrement function $d$ to indicate wrapping around
the circle $p$ times.  Hence the equation of left boundary
of the top link of $A_{p/q}$ for $q>1$ is given schematically by
$$  f_2 \cdots f_2 d\,f_2 \cdots f_2 d\,f_2 (t_1) = t_1,
    {\it ~~~~~~~~~~~~(C4)}
$$
where the left side of (C4) is a composite of $q$
copies of $f_2$ and $p$ copies of $d$.  The positions of
the $d$'s are given by the following rule:  Counting from
right to left in (C4), we insert a ``$d$'' to the
left of the $i$th $f_2$ for each $i$ such that
$$
   0 < i p \pmod{q} \leq p {\it ~~~~~~} (i=1, \dots, q-1).
$$
For the case $q=1$ the left boundary equation is simply
$f_2(t_1) = t_1$, which is easily solved for $\Omega$ to
obtain

$$\Omega  = -(S-1)t_1 = \Omega_{\rm left}(S,0/1,\delta).$$

Equation (C4) contains the quantities $\Omega$, $S$,
$\delta$ [recall $t_1=(1-\delta)/2$], $p$, and $q$.
By using properties of the simple functions $f_2$ and $d$ it is
relatively straightforward to solve the equation (C4) for
$\Omega$, with the following result:

\medskip \noindent
{\it Left tip boundary above the top node:}

\noindent For $q>1$ and $p/q$ in lowest terms,
$$
   \Omega = \Omega_{{\rm left}}(S,p/q,\delta)
   =  (S-1) [{ {N_{\rm left}(S,p/q)} \over {S^q -1}} - t_1 ],
   {\it ~~~~~~~~(C5)}
$$
where
$$
   N_{\rm left}(S,p/q) =
   \sum^{q-1}_{i=1 \atop 0 < ip \pmod{q} \leq p} S^{q-i}.
$$

The right boundary of $A_{p/q}$ can be obtained in a similar
manner to the above.  On the right boundary, a $p/q$-cycle of
the tip map contains the other turning point,
$t_0 = -(1-\delta)/2$, which leads to:
$$
d\, f_2 \cdots f_2 d\,f_2 \cdots f_2 d\,f_2 \cdots f_2(t_0)=t_0,
    {\it ~~~~~~~~~~~~(C6)}
$$
where the left side of (C6) is a composite of $q$
copies of $f_2$ and $p$ copies of $d$., and where,
counting from
right to left in (C6), we insert a ``$d$'' to the
left of the $i$th $f_2$ for each $i$ such that
$$
   0 \leq ip \pmod{q} < p  {\it ~~~~~~} (i=2, \dots, q).
$$

Solving equation (C6) for $\Omega$, again after some
straightforward but nontrivial computations, gives the
equation for the right boundary of $A_{p/q}$ for the tip
family above the top node:

\medskip \noindent
{\it Right tip boundary above the top node:}

\noindent  For $q>1$ and $p/q$ in lowest terms,
$$
   \Omega = \Omega_{\rm right}(S,p/q,\delta)
   =  (S-1) [{ {N_{\rm right}(S,p/q)} \over {S^q -1} }- t_0 ],
  {\it ~~~~~~(C7)}
$$
where,
$$
   N_{\rm right}(S,p/q) =
   \sum^{q}_{i=2 \atop 0 \leq ip \pmod{q} < p} S^{q-i}
   = N_{\rm left}(S,p/q) + 1 - S^{q-1}.
$$

The boundary equations (C5) and (C7) are valid for the tip
family version of $A_{p/q}$ anywhere above the top node,
including on and above the critical line.  By subtracting
equation (C5) from (C7), for example, one obtains the exact
equation for the length of the $p/q$-stability intervals
on the critical line, which was given in Section 4 of this
paper and used in the proof of Theorem 5:

\medskip \noindent
{\it Length of p/q-stability interval on the critical line:}
$$
|L^{\circ}_{p/q}| = \Omega_{\rm right} - \Omega_{\rm left} =
    { (S-1)^2  \over S(S^q-1)},  {\it ~~~~~~~~(C8)}
$$
where on the critical line, $S=S_c=1/(1-\delta)$.
Equation (C8) for the width of $A_{p/q}$
for the tip family, at height $S$ in the parameter plane,
is valid for any $S$ above the top node
of $A_{p/q}$, including on and above the critical line.

When $q=1$, $A_{p/q}$ has no nodes, since then
$\lceil q \delta \rceil - 1 = 0$, and hence the complete left or
right boundary of a 0/1 or 1/1 tip
family tongue is easily found in a manner similar to the
way we found the
left boundary $\Omega = \Omega_{\rm left}(S,0/1,\delta)$.
The resulting 0/1 and 1/1 boundary equations are:

\medskip \noindent
{\it Complete tip family boundaries for q=1:}

\medskip
$ \Omega = \Omega_{\rm left}(S,0/1,\delta) \ \,= -(S-1)t_1$,

$ \Omega = \Omega_{\rm right}(S,0/1,\delta) = -(S-1)t_0$,
                                   {\it \hskip1.5in       (C9)}

$ \Omega = \Omega_{\rm left}(S,1/1,\delta) \ \,= -(S-1)t_1 + 1$,

$ \Omega = \Omega_{\rm right}(S,1/1,\delta) = -(S-1)t_0 + 1$,

\medskip \noindent
where $t_1 = (1-\delta)/2$ and $t_0 = -t_1$.

\medskip \noindent
Note that each of these lines in the $(\Omega,S)$ plane has
a slope of $\pm 2/(1-\delta)$ when $S$ is expressed as
a function of $\Omega$.

\bigskip       
{\bf C-1.4.  Boundaries of $A_{p/q}$ below the top node.}

\medskip
In this subsection we briefly describe the exact boundary
equations for $A_{p/q}$ below the top node.

For a given $S>1$ it is easy to determine which link of the
$p/q$-sausage is cut by the horizontal line at $S$:  just
test the left side of equation (C1) for
$k=1, 2, \dots, \lceil q\delta \rceil -1$; the
first $k$ that gives a value $\leq 0$ tells us that the
height $S$ corresponds to the $k$th link from the top,
where the part above the top node is link 0.

For every one of the sausage links one can write the left and
right boundary equations
$$
t_{\Omega,S;\delta}^q(t_i) = t_i, ~~~(i=1,0 ~{\it respectively,})
$$
in forms similar to (C4) and (C6).  The fact that
on the boundary of link $k$,
$k$ of the $p/q$-cycle members have escaped from the
big slope region of the circle map $t_{\Omega,S;\delta}$, means
that $k$ of the ``$f_2$''s in (C4) and (C6) must be replaced by
``$f_1$''s or ``$f_3$''s.  It is straightforward to replace
(C4) and (C6) with versions that work for the boundaries of
the $k$th link from the top for arbitrary $k$ in the admissible
range $0, 1, \dots, \lceil q\delta \rceil -1$.  We will not
dwell on the equations for arbitrary $k$, but will simply
illustrate with the following example.

\medskip \noindent
{\it Example:}

\noindent  The right boundary equations for the 3 links
of $A_{2/5}$, are:
$$ k=0: ~~~ d\, f_2 f_2 d\, f_2 f_2 f_2 (t_0) = t_0$$
$$ k=1: ~~~ d\, f_2 f_2 d\, f_3 f_2 f_2 (t_0) = t_0$$
$$ k=2: ~~~ d\, f_3 f_2 d\, f_3 f_2 f_2 (t_0) = t_0.$$
By writing the specific formulas of each of the ``$f$''
and ``$d$'' functions, one can solve each of these
equations explicitly for $\Omega$ in
terms of $S$ and $\delta$.  For example, the middle equation
above, when solved for $\Omega$, gives the right boundary
of link 1:
$$
  \Omega = \Omega_{\rm right} = {
  {\delta (1+\delta) +   (1+\delta)S^2 - (1-\delta)S^3 +
(1-\delta)S^4 - (1-\delta)^2 S^5}
               \over
  { 2[ \delta + \delta S + (1+\delta)S^2 + \delta S^3 -
(1-\delta)S^4]}
}.
$$

For the purpose of generating computer-drawn curves of the
left and right boundaries of the tip family
tongues $A_{p/q}$, we have
written a general computer algorithm that for a given
$S \geq 1$, solves the left and right boundary equations
corresponding to (C4) and (C6) for the appropriate link.  The
algorithm finds the exact (up to computer accuracy) values of
$\Omega$ on the left and right boundaries for a given $S$ by
recursively computing the left side of the equation as a
affine function of $\Omega$ and then solving the
affine equation for $\Omega$.  This computer algorithm has
been used to generate accurate drawings of the left and right boundaries of
many of the tongues in this paper, including parts of the
boundaries of the stability regions shown later in Appendix C.

\bigskip   
{\bf C-1.5.  The unbounded part of $A_{p/q}$.}

\medskip
   As explained in Section C-1.3, equations (C5) and (C7) give
the left and right boundaries, respectively, of the tip
family $A_{p/q}$ above the top node, which also includes
the portion on and above the critical line.  So we have nothing
more to say about the left and right boundaries above the
line, except to note that in equations (C5) and (C7) the two
numerators $N_{\rm left}$ and $N_{\rm right}$ each are
polynomials in $S$ of degree less than $q$.  Hence equations
(C5) and (C7) imply that for large $S$, the right and left
boundaries of $A^+_{p/q}$ in the $(\Omega,S)$ plane are
asymptotic to straight lines with slope $\pm 2/(1-\delta)$.
Thus, all tongue boundaries are asymptotically parallel as $S
\rightarrow \infty$.

\medskip 
{\bf C-1.6.   Boundaries of other $p/q$ regions above
the critical line.}

\medskip
   By considering the slopes of a tip map $t_{\Omega,S;\delta}$
and its $q$th iterate $t^q$, it is fairly straightforward
to deduce other statements concerning the exact borders
of the various regions previously
discussed in this paper.  These
properties are summarized below, where for fixed $\delta$,
$A^{stab}_{p/q}$ = the $p/q$-stability region,
$A^{stab'}_{p/q}$ = the extended stability region where
$t_{\Omega,S;\delta}$ has either a stable $p/q$-cycle or a
$p/q$-ordered chaotic attractor (as defined just before
Theorem 18), and $L_{p/q}$ = the region where
$t_{\Omega,S;\delta}$ has the unique rotation number $p/q$.
Some of these regions were shown in Figures 7, 13, 14, 15
and 18.  The boundaries in those figures were obtained
from the equations in this appendix.

\medskip \noindent
{\it The heights of the regions:}

\noindent
Let $S = $
$S_{\rm max}(q)$, $S'_{\rm max}(q)$, and $S_{\rm tip}(q)$
be the heights in the $(\Omega,S)$ plane of
the regions $A^{stab}_{p/q}$,
$A^{stab'}_{p/q}$, and $L_{p/q}$, respectively.  Then:

\medskip
    {\it a)~~} The three heights satisfy, respectively, the
following three polynomial equations:

\medskip
$  S_{\rm max}:~~~  (1-\delta)S^q - S^{q-1} - \delta = 0, $

$  S'_{\rm max}:~~~ (1-\delta)S^{q+1} - S^q - S + 1 = 0, $

$  S_{\rm tip}:~~~~  (1-\delta)S^{q+1} - S^q - (2-\delta)S
                       + 2 = 0. $

\medskip
   {\it b)~~}
$ S_{\rm tip} > S'_{\rm max}  > S_{\rm max} > S_c$.

\medskip
   {\it c)~~}
All three heights decrease with $q$ and as
$q \rightarrow \infty$ they approach the critical line
$S = S_c = 1/(1-\delta)$ exponentially fast in q.

\medskip \noindent
Of course, with the help of Newton's method or other numerical
procedures, the above three heights can easily be computed
very accurately and quickly.

\medskip \noindent
{\it The top boundary of $L_{p/q}$:} 

\noindent
The left (respectively, right) boundary of $L_{p/q}$
above the critical line, is the limit of the right (left)
boundaries of the neighboring $A_{\omega}$
as $\omega \rightarrow p/q$ from the left (right).
Since in practice these
neighboring boundaries converge rather quickly with $\omega$,
one can use the results of Section C-1.3 to draw quite
accurate boundaries of $L_{p/q}$ above the critical line.
The left and right boundaries of $L^+_{p/q}$ intersect at
height $S = S_{\rm tip}(q)$, so that the top of $L_{p/q}$
is somewhat ``ice cream cone'' shaped, as is seen in Figure 7.
Of course, below the line, $L_{p/q}$ is the same as $A_{p/q}$.

\medskip \noindent
{\it Top boundaries of $A^{stab}_{p/q}$ and  
  $A^{stab'}_{p/q}$:}

\noindent
$A^{stab}_{p/q}$ is the portion of $A_{p/q}$ bounded above
by the horizontal line $S = S_{\rm max}$ described above.
Similarly,
$A^{stab'}_{p/q}$ is the portion of $A_{p/q}$ bounded above
by the horizontal line $S = S'_{\rm max}$.

\medskip \noindent  
{\it Boundary of the p/q-topologically regular region:}

\noindent
{}From Section 9 of this paper, the part of $A_{p/q}$ that has
zero topological entropy is $A^{stab}_{p/q} \cap L_{p/q}$, so
this region can be accurately plotted by using boundaries
determined in the above discussions.

\medskip
With the above properties we can draw the various tip family
$p/q$-regions very accurately along with their unions and
intersections.  These properties were used to generate Figures
C3 to C15 showing the exact boundaries of $A^{stab}_{p/q}$
as $\delta$ increases from 0.1 to 0.9999.  These figures
illustrate many of the concepts discussed in this paper.

\bigskip
{\bf C-1.7.  The zipper theorems.} 

\medskip
We call Theorems 17 and 18 the {\it Zipper Theorems} because
of the way they indicate that the Farey neighbors
of $A^{stab}_{p/q}$ and $A^{stab'}_{p/q}$ unzip
from each of them  as $\delta$ decreases.  These theorems
are represented accurately in Figures 15abc and 18 of the
main body of this paper, using the boundary equations and
algorithms described above.

\bigskip           
{\bf C-1.8.  Monte Carlo plotting of $A^{stab}_{p/q}$.}

\medskip
One can also obtain reasonably accurate plots of the
bifurcation structure of our families by using a numerical
Monte Carlo algorithm such as the following, where color
$C(p,q)$ is a predefined color table [normally $C(p,q)=q$]:

%

\medskip
\goodbreak 
{\it
$\cdot$~~~~For each $(\Omega,U)$ in a rectangle, do:

$\cdot$~~~~~~~~Choose x randomly

$\cdot$~~~~~~~~Repeat x=f(x) M times

$\cdot$~~~~~~~~For q=1 to $q_{max}$

$\cdot$~~~~~~~~~~~~If $|f^q(x)-x| < \epsilon$ then

$\cdot$~~~~~~~~~~~~~~~~Find the winding number p/q

$\cdot$~~~~~~~~~~~~~~~~Plot the point $(\Omega,U)$ in color
                       $C(p,q)$

$\cdot$~~~~~~~~~~~~~~~~Exit the q loop

$\cdot$~~~~~~~~~~~~End if

$\cdot$~~~~~~~~End the q loop

$\cdot$~~~~End the do loop
}

\medskip \noindent
This algorithm was used to plot Figure 10 for the plateau
family regions $A^{stab}_{p/q}$, $q \leq 5$, $\delta=1/2$.
It was also used for Figure C17 ($A^{stab}_{p/q}$ for the
tip family, $q \leq 7$, $\delta=1/2$) which should be compared
with the exact version in Figure C7.  The Monte Carlo
algorithm gives good approximations of all the exact versions
in Figures C3-C15.

\bigskip \noindent   
{\bf C-2.  Plateau Family Boundaries above the Critical Line.}

\medskip  
In this section we will describe the boundaries of the
various regions above the critical line for the
plateau family.

Some of the $p/q$ regions for the plateau family are
considerably more complicated than for the tip family, because
of the flat spots (plateaus) on the graph of the lift
$P_{\Omega,H;\delta}$ and its iterates.  These flat spots
allow infinite sequences of period doubling, tripling, etc.,
as is the case with the standard sine family and the logistics
family.

We do not have exact descriptions of the upper
boundaries for the plateau stability
and topologically regular $p/q$-regions, as we did for the
tip family.  However some of the other regions are actually
simpler to describe exactly than for the tip family and
can be obtained by elementary methods.  To understand fully
the structure of all parts of the stability regions, one needs
to consult works on symbolic dynamics, such as [MaT1-3].

Following is a summary of our results about the plateau
family above the critical line.  The formulas below come in
an elementary manner from the slopes in the plateau maps
($\pm S_c$ and $0$) and from our exact equations in Section C-1
about the boundaries of the $p/q$-intervals on the critical
line.

\bigskip 
Assume $\delta$ is fixed, $0 < \delta < 1$ and that
$p/q$ is in lowest terms.

\medskip \noindent
{\it Top Boundary of $A_{p/q}$:}  

\noindent
The top of $A_{p/q}$ for the plateau family in the
$(\Omega, H)$ plane is the horizontal line
    $$  H = H_{\rm max} = { \delta \over 4(1-\delta) }. $$

\noindent
{\it Side boundaries of  $A_{p/q}$:} 

\noindent
The left and right boundaries of  $A_{p/q}$ are the
straight lines:
$$ H = H_{\rm left}(\Omega) = (\Omega_L - \Omega) / \delta, $$
$$ H = H_{\rm right}(\Omega) = (\Omega - \Omega_R) / \delta, $$
where $(\Omega_L, \Omega_R)$ is the $p/q$-stability interval
on the critical line for the tip family.
$\Omega_L$ and $\Omega_R$ are given exactly
by (C5) and (C7) in Section C-1 with $S=S_c=1/(1-\delta)$.

\medskip
The next result follows easily by taking a limit of boundaries
of the neighboring regions.  All of these boundaries have
slope $\pm 1/ \delta$, from the previous property.

\medskip \noindent
{\it The $L_{p/q}$ cone:} 

\noindent
For the plateau family, $L^+_{p/q}$ is shaped like an
inverted cone (or a tent) with
straight line sides having the equations:
$$  H = (\Omega - \Omega_L) / \delta {\it ~~~~(left)}, $$
$$  H = (\Omega_R - \Omega) / \delta {\it ~~~~(right).} $$
When $\delta < 1 - 3^{-1/q}$, the cone is truncated at
height $H = H_{\rm max}$.

\medskip \noindent
{\it The stability region $A^{+,stab}_{p/q}$:} 

\noindent
For the plateau family, the stability region $A^{+,stab}_{p/q}$
is composed of several disjoint parts, as follows.  These
regions are shown in the numerically generated Figure 10
and schematically in Figure C18.

\medskip
    {\it a)~~}  a lower, winged-shaped part with bottom
    boundary $H=0$, left and right boundaries the same
    as $A_{p/q}$ above, and top boundary
    composed of the two lines

    $$
    H = \pm (\Omega - \Omega_{\rm mid})  {  H_{\rm max} \over
        (\Omega_R - \Omega_L)/2  +  \delta H_{\rm max} }
    $$
    where $\Omega_{\rm mid} = (\Omega_L + \Omega_R)/2$,

\medskip
   {\it b)~~}  an infinite number of swallow-like period
multiplying regions with rotation numbers $(kp)/(kq)$ for
$k = 2,3,\dots$, which lie above the ``wing'' described in
a) above.

\medskip
    {\it c)~~} a finite number of swallow-like $p/q$-regions that
    lie outside $L^+_{p/q}$ and outside the period multiplying
    region.  See Figure 10 for a 2/5-swallow of this type.

\medskip \noindent
{\it Location of swallow tips:} 

\noindent
All of the above wing and swallow-like $(kp)/(kq)$-regions
have the tips of their wings and tails on the line
$H = H_{\rm max}$.

\medskip
As we noted earlier, understanding the structure of these
swallows requires knowledge of the topology of symbol
sequence dynamics but is fairly well understood.
In [MaT1-3] it is explained how the boundary of topological
chaos is squeezed between the regions of period $2^k q$
$(k = 1,2,\dots)$ and those of period $m2^k q$ where $m$
is odd.  As a result the $p/q$-region of topological regularity
includes the cone $L^+_{p/q}$ chopped off on the top in a
(fractal) jagged way somewhat as suggested by Figure C19.

Note that the regions for the plateau family,
above the critical line, preserve most of the topological
features of the regions for the classical sine circle
family.  Compare, for example, the topological structures
of $p/q$-regions for the plateau family in Figure 10
with those for the sine family in Figure C20.  (Both
figures were generated numerically.)  One can see that
for each structure that appears in Figure 10 a
corresponding, similar but distorted, version appears in
Figure C20, although those in Figure C20 may be difficult
to see because the sine $p/q$-wings and swallows appear
to shrink faster with $q$ than do the plateau versions.
Also note that the plateau wings and swallows are bounded,
while the corresponding sine versions have infinitely long
wings and tails.
\filbreak
\medskip
\bigskip \noindent
{\bf Appendix D:~~  Measures below the Critical Line.}

\bigskip
A natural question to ask is, for $1 < S_0 < S_c = 1/(1-\delta)$,
what is the one-dimensional Lebesgue measure of the intersection
of the horizontal line $S=S_0$ with the union of the
$A_{p/q}$,
$0 \le p/q \le 1$?  Since the  ``sausages'' do not intersect
each other below the critical line, the sum of these interval
lengths on the line $S=S_0$, for $0 \le \Omega \le 1$, is no
larger than 1.  But is it less than 1 or is it equal to 1?
As was remarked at the end of Section 7,
for ``nice'' smooth families of circle maps,
including the standard sine family, this measure is known
[He1] to be less than 1.  But for certain families of
maps that are continuous on the circle and $C^2$ smooth except
at one point, the measure is known to be equal to 1 [VK],
even below the critical line: {\it i.e.}, the set of $\Omega$ corresponding
to irrational winding
numbers has zero measure for any {\it non-zero} value of the
nonlinearity parameter $S-1$.  Our maps
resemble those of [VK] but have {\it two}
points of nonsmoothness (rather than one point),
 namely, the {\it turning points}, $(1 \pm \delta)/2$.
Based on the numerical results we describe
below, we believe that {\it for our family below the
critical line the measure of the rational (``mode-locked'') intervals
is less than 1}.

Using the algorithm described in Section C-1
of the Appendix, we have written
a computer code that, for given
values of $\delta$, $S=S_0<S_c$ and
$q_{\rm max}$, finds the ``exact'' (up to computer accuracy)
endpoints of the intersection
of the line $S=S_0$ with all $A_{p/q}$ for $q =$
$1, \dots , q_{\rm max}$ and $p = $ $0,1, \dots, q$ where
$p$ and $q$ are relatively prime.  The program then sums the
lengths of these intervals on the line $S=S_0$.
The resulting approximate measures for various values of $S_0$,
$1 \le S_0 \le S_c$ $= 1/(1-\delta)$, are shown in Figure D1.
The measures in Figure D1 are a result of summing as many
as 12,232 interval lengths ($q_{\rm max} = 200$) for each
value of S, using 29 significant decimal digits of precision.

It should be emphasized that the results in Figure D1
are preliminary numerical results and are not
conclusive evidence that the measure of the mode locked
intervals below the critical line are less than 1.  More
rigorous work is needed on this question.
%
%

\bigskip

{\bf REFERENCES.}
\bigskip
\frenchspacing

\item{[AKM]}  R.L. Adler, A.G. Konheim, and M.H. Mc Andrew,
``Topological entropy",   {\sl        Trans. Amer. Math. Soc. \bf 114}
(1965) 309--319.
\item{[AM1]} L. Alsed\`{a} and F. Man\~{o}sas, ``Kneading theory and
rotation intervals for a class of circle maps of degree one", {\sl
Nonlinearity \bf 3} (1990) 413--452.
\item{[AM2]} L. Alsed\`{a} and F. Man\~{o}sas, ``The monotonicity of
the entropy for a family of degree one circle maps",
Preprint Barcelona (1992).
\item{[Ar]} V.I. Arnold, ``Small denominators I, Mappings of
the circle onto itself", {\sl Izv. Akad. Nauk SSSR Ser. Mat.\bf  25}
(1961) 21--86, (English translation {\sl Transl. Amer. Math. Soc.
\bf  46} (1965) 213--284).
\item{[Ar2]}V. I. Arnold, ``Cardiac arrhythmias and circle mappings,'' {\it
Chaos}
{\bf 1} (1991) 20-25; this article is a previously unpublished section
of Arnold's 1959 diploma dissertation which was omitted in the published
reference [Ar].
\item{[Bak]} P. Bak, ``The Devil's Staircase,'' {\it Physics Today},
(Dec. 1986) 38-45.
\item{[BBr]}, P. Bak and R. Bruinsma, ``One-dimensional Ising model and the
complete devil's staircase,'', {\it Phys. Rev. Lett.} {\bf 49} (1982) 249-251.
\item{[Be]}  J. Bernoulli, ``Sur une nouvelle esp{\`e}ce de calcul", {\sl
Recueil pour les Astronomes {\bf 1} (Berlin)} (1772) 255--284.
\item{[BJa]} P.M. Blecher and M.V. Jakobson,``Absolutely continuous
invariant measures for some maps of the circle", {\sl in} {\bf Statistical
Physics and Dynamical Systems}
({\sl Birkh\"auser, Boston. Basel. Stuttgart}, 1985).
\item{[BlF]} L. Block and J. Franke, ``Existence of periodic points for
maps of ${\bf S}^1$, " {\sl Inv. Math. \bf 22} (1973) 69--73.
\item{[BlGMY]} L. Block, J. Guckenheimer, M. Misiurewicz, and L.S. Young,
``Periodic  points and
topological entropy of one dimensional maps",  in {\bf
Springer Lecture  Notes in Math. Vol. 819} ({\sl Springer, Berlin}, 1980).
\item{[Boh]} P. Bohl, ``Uber die hinsichtlich der unabh\"angigen variabeln
periodische Differentialgleichung erster Ordnung", {\sl Acta Math. \bf 40}
(1916) 321--336.
\item{[BBJ]} T. Bohr, P. Bak, and M. H. Jensen, ``Transition to Chaos
by interaction of resonances in dissipative systems.
II. Josephson junctions, charge-density waves, and standard maps.''
{\it Phys. Rev. A} {\bf 30} (1984) 1970-1981.
\item{[Bow]} R. Bowen,``Entropy for group endomorphisms and homogeneous
spaces", {\sl Trans. Amer.
Math. Soc. \bf 153} (1971) 401--414.
\item{[BFr]} R. Bowen and J. Franks, ``The periodic points of maps of the disk
and the interval", {\sl
Topology \bf 15} (1976) 337--342 .
\item{[Boyd]} C. Boyd, ``On the structure of the family of Cherry fields on the
torus,''  {\sl Ergod. Th. \&
Dynam. Sys. \bf 5} (1985) 27--46.
\item{[Boyl]} P.L. Boyland, ``Bifurcations of circle maps: Arnol'd tongues,
bistability and rotation
intervals," {\sl Commun. Math. Phys. \bf 106} (1986) 353--381.
\item{[BMT]} K.M. Brucks, M Misiurewicz and C. Tresser, ``Monotonicity
properties  of the family of trapezoidal maps," {\sl Commun. Math. Phys. \bf
137} (1991) 1--12.

\item{[BT]} K.M.  Brucks and C. Tresser, ``A Farey
tree organization of locking regions for simple circle maps," to appear in
{\it Prog. Amer. Math. Soc.}
\item{[BrB]} R. Bruinsma and P. Bak., ``Self-similarity and fractal dimension
of
the Devil's staircase in the one dimensional Ising model,''
{\it Phys. Rev.}, {\bf B27} (1983) 5824-5825.
\item{[CGT]} A. Chenciner, J.M. Gambaudo and C. Tresser, ``Une remarque sur la
structure des
endomorphismes de degr\'e un du cercle," {\sl C.R. Acad. Sc. Paris \bf t.299
s\'erie I} (1984) 771--773.
\item{[Ch]}  E.B. Christoffel, ``Observatio Arithmetica," {\sl Annali di
Mathematica, 2nd series \bf 6} (1875) 148--152.
\item {[CTA]} P. Coullet, C. Tresser, and A. Arn{\'e}odo, ``Transition to
turbulence for  doubly periodic flows," {\sl Phys. Lett. \bf 77A}
(1980) 327--331.
\item{[DGMT]} S. P. Dawson, R. Galeeva, J. Milnor and C. Tresser, ``A
monotonicity conjecture for real cubic maps," {\sl To appear in Proceedings of
the NATO Advanced Study
Institute on Real and Complex Dynamical Systems, Hiller{\o}d, June 1993.} ,
IMS-SUNY Stonybrook
preprint
series, \# 1993/11 (1993).
\item {[De]} A. Denjoy, ``Sur les courbes d\'efinies par les \'equations
diff\'erentielles \`a la surface du tore," {\sl J. de Math. Pures et
Appl. \bf 11} (1932) 333--375.
\item {[EKT]} A. Epstein, L. Keen and C. Tresser, `` The set of maps
$F_{a,b}:x \mapsto x+a+{b\over 2\pi} sin(2\pi x)$ with any given
rotation interval is contractible," to appear in {\it Comm. Math. Phys.}
\item{[FKS]} M. J. Feigenbaum, L. P. Kadanoff, and S. J. Shenker,
``Quasiperiodicity
in Dissipative Systems: A Renormalization Group Analysis,'' {\it Physica D}
{\bf 5}
(1982) 370-386.
\item{[FT]} B. Friedman and C. Tresser, ``Comb structure in hairy
boundaries: some transition problems for circle maps," {\sl Phys. Lett. A \bf
117} (1986) 15--22.
\item{[Ga]} R. Galeeva, ``Kneading sequences of
piecewise linear bimodal maps," {\sl to appear in Chaos}.
\item{[Ga2]} R. Galeeva, ```House Maps' of the Interval,'' unpublished.
\item {[GMT]} R. Galeeva, M. Martens, and C. Tresser, ``Inducing, slopes, and
conjugacies," IMS-SUNY
Stonybrook preprint series, \# 1994/3 (1994).
\item {[GaT]} R. Galeeva and C. Tresser, ``Piecewise linear
discontinuous double  coverings of the circle," {\sl Proc. Amer
Math. Soc. \bf 118} (1993) 285--291.
\item {[GTr]} J.M. Gambaudo and C. Tresser, ``Transition vers le chaos pour les
applications de degr{\'e}
un du cercle," in {\bf ``le Chaos"}, S{\'e}rie
Synth{\`e}ses, ({\sl Editions  Eyrolles, Paris}, 1988).
\item{[GB]} L. Glass and J. B\'{e}lair, ``Continuation of Arnold tongues
in mathematical models of periodically forced biological
oscillations,'' {\it Nonlinear Oscillations in Biology and
Chemistry}, Lecture Notes in Biomathematics, edited by H. G.
Othmer, (Springer-Verlag, Berlin, 1986) 232-243.
\item{[Gl]} L. Glass,  ``Cardiac arrhythmias and circle maps - A classical
problem,'' {\it Chaos} {\bf  1} (1991) 13-19.
\item {[GM]} L. Glass and M. C. Mackey,  {\it From Clocks to Chaos, The
Rhythms of Life}, Princeton Univ. Press (Princeton, 1989).
\item{[GoT]} L. Goldberg and C. Tresser, ``Rotation orbits and the Farey
tree," {\sl Preprint I.B.M.} (1991).
\item{[HW]} G.H. Hardy and E.M. Wright, ``{\bf An
Introduction to the Theory of Numbers},"  ({\sl Clarendon,
Oxford}, 1979).
\item{[He1]} M.R. Herman, ``Mesure de Lebesgue et nombre de rotation,''
in {\bf Springer Lecture  Notes
in Math. Vol.  597} ({\sl Springer, Berlin,} 1977). 271-293.
\item{[He2]} M.R. Herman, `` Sur la conjugaison diff{\'e}rentiable des
diff{\'e}omorphismes {\`a} des rotations," {\sl Pub. Math. I.H.E.S. \bf 49}
(1980) 5--234.
\item{[He3]} M.R. Herman, ``Majoration du nombre de cycles p\'{e}riodiques pour
 certaines familles  de
diff\'{e}omorphismes du cercle,'' {\sl An. Acad.
Brasil. Ci\^{e}nc. \bf 57} (1985) 261--263.
\item {[HC]} D. Horton and D. K. Campbell, ``Partial Period Doubling
Sequences and Symbolic Dynamics in `House Maps' of the Interval,'' to be
published.
\item{[Hu]} C. Huyghens, letter to his father, dated February 26, 1665.
{\it Ouevres completes des Christian Huyghens}, Vol. 5, p. 243
M. Nijhoff, ed.,(Societ\'e Hollandaise des Sciences, The
Hague, Netherlands (1893). (see references [Bak], [StSt] for a discussion of
this
observation).
\item{[JBB1]} M. Jensen, P. Bak, and T. Bohr, ``Complete devil's staircase,
fractal  dimension  and
universality of mode-locking structure in the circle map," {\sl Phys. Rev.
Letters \bf 50}  (1983) 1637--
1639.
\item{[JBB2]} M. Jensen, P. Bak, and T. Bohr, ``Transition to chaos by
interaction of resonances in dissipative  systems. I. Circle maps,'' {\sl
Phys. Rev. A \bf 30} (1984) 1960--1969.
\item{[Kh]} K.M. Khanin, ``Universal estimates for critical circle
mappings," {\sl Chaos \bf 1} (1991) 181--186.
\item{[MaT1]} R.S. Mackay and C. Tresser, ``Transition to topological
chaos for circle maps," {\sl Physica \bf 19D} (1986) 206--237 \& {\bf 29D}
(1988) 427.
\item {[MaT2]} R.S. Mackay and C. Tresser, ``Some flesh on the skeleton: the
bifurcation  structure of
bimodal maps," {\sl  Physica  \bf 27D} (1987) 412--
422.
\item {[MaT3]} R.S. Mackay and C. Tresser, ``Boundary of chaos for bimodal maps
of the  interval," {\sl
J. London Math. Soc. \bf 37} (1988) 164--181.
\item {[Ma]}  A.A. Markoff, ``Sur une question de Jean
Bernoulli," {\sl Math. Ann. \bf  19} (1882) 27--36.
\item {[MTr]} M. Martens and C. Tresser, ``Forcing in $C^0([0,1])$ and
renormalization in $PL([0,1])$," to appear.
\item{[MSE]} R. Mainieri, T. S. Sullivan, and R. E. Ecke, ``Two-Parameter
Study of the Quasi-periodic Route to Chaos in Convecting $^3He$-
Superfluid $^4 He$ Mixtures'', {\it Phys. Rev. Lett.} {\bf 63} (1989)
2357-2360.
\item{[MT]} M. Martens and C. Tresser, ``Forcing of Periodic Orbits
for Interval Maps and Renormalizations of Piecewise Affine Maps,''
SUNY StonyBrook, Institute for Mathematical Sciences, preprint \# 1994/17.
\item{[MM]} S. Martin and W. Martienssen, ``Circle Maps and Mode Locking in
the Driven Electrical Conductivity of Barium Sodium Niobate Crystals,''
{\it Phys. Rev. Lett.} {\bf 56} (1986) 1522-1525
\item{[MS]} J. Masleko and H. L. Swinney, ``A Complex Transition Sequence in
the
Belusov-Zhabotinskii Reaction,'' {\it Phys. Scr.} {\bf T9}, (1985) 35-39.
\item {[Me]}  W. de Melo, ``Full families of circle maps," to appear.
\item {[MSS]} N. Metropolis, M.L. Stein, and P.R. Stein, ``On finite limit sets
for transformations on the
unit interval," {\sl{J. Comb.
Theory}} {\bf{15}} (1973) 25--44.
\item {[MSt]}  W. de Melo and S. van Strien, {\bf {One Dimensional
Dynamics}} (Ergebnisse der Mathematik und ihrer Grenzgebiete. 3. Folge, Vol.
25) ({\sl Springer-
Verlag, Berlin}, 1993).
\item{[MTh]} J.Milnor and W. Thurston, ``On iterated maps of the interval,"  in
{\bf Springer Lecture
Notes in Math. Vol.  1342} ({\sl Springer, Berlin,} 1988).
\item{[MH]} M. Morse and G.A. Hedlund, ``Symbolic dynamics II: Sturmian
trajectories," {\sl Amer. J.
Math. \bf  62} (1940) 1--42.
\item{[Mi1]} M. Misiurewicz, ``Horseshoes for mappings of the interval," {\sl
Bull. Acad.  Pol.  Ser. Sci.
Math. \bf 27} (1979) 167--169.
\item{[Mi2]} M. Misiurewicz, ``Twist sets for maps of the circle," {\sl Erg.
Th. \& Dyn. Syst. \bf 4}
(1984) 391--404.
\item{[MSz]} M. Misiurewicz and W. Szlenk, ``Entropy of piecewise monotone
maps," {\sl Studia  Math.
\bf 67} (1980) 45--68.
\item{[MV]} M. Misiurewicz and E. Visinescu, ```Kneading sequences of skew tent
maps," {\sl Ann.Inst.
H.Poincar\'e \bf 27} (1991) 125--140.
\item {[Mu]} P. Mumbr{\'u}, {\bf{Estructura
Peri{\`o}dica i Entropia Topol{\`o}gica de les
Aplicacions Bimodals}} Ph. D. Universitat
Aut{\`o}noma de Barcelona (1987).
\item{[NPT]} S. Newhouse, J. Palis and F. Takens, ``Stable
families of diffeomorphisms," {\sl  Pub. Math. I.H.E.S. \bf 57}
(1983) 5--71.
\item{[ORSS]} S. Ostlund, D. Rand, J. Sethna, and E. D. Siggia, ``Universal
Properties of the Transition
from Quasi-Periodicity to Chaos in Dissipative Systems," {\sl Physica \bf 8D}
(1983) 303--342.
\item{[PG]}R. Perez and L. Glass, ``Bistability, Period Doubling
Bifurcations and Chaos in a Periodically Forced Oscillator,''
{\it Phys. Lett. A} {\bf 90} (1982) 441-443.
\item{[Po]} H. Poincar\'e, ``Sur les courbes d\'efinies par des \'equations
diff\'erentielles," {\sl
J.Math.Pures et Appl. $4^{eme}$ s\'erie,  \bf 1} (1885) 167--244. Also in
``{\bf Oeuvres Compl\`etes, t.1}"
({\sl
Gauthier-Villars, Paris}, 1951).
\item {[STZ]} R. Siegel, C. Tresser, and G. Zettler, ``A decoding problem in
dynamics  and in number
theory," {\sl Chaos \bf 2} (1992) 473--493.
\item{[Sm]}  H.J.S. Smith, ``Note on continued fractions,"
{\sl The Messenger of Mathematics \bf VI} (1877) 1--14.
\item { [SHL]} J.Stavans, F. Heslot, and A. Libchaber, ``Fixed Winding
Number and the Quasiperiodic Route to Chaos in a Convective Fluid,''
{\it Phys. Rev. Lett.} {\bf 55} (1985).x=
\item{[St]} S. H. Strogatz, ``Norbert Wiener's Brain Waves,'' {\it Lecture
Notes in Biomathematics, Vol. 100,} (Springer, 1994) 122-138.

\item{[StSt]} S. H. Strogatz and I. Stewart, ``Coupled Oscillators and
Biological
Synchronization, {\it Scientific American} (1993) 102-109.
\item{[Sw]} G. \'{S}wi\c{a}tek, ``Rational rotation numbers for maps of the
circle,''  {\sl Commun. Math.
Phys. \bf 119} (1988) 109--128.
\item{[Tri]} C. Tricot, ``Douze d\'efinitions de la densit\'e
logarithmique," {\sl C.R. Acad. Sc. Paris \bf t.293 s\'erie I} (1981)
549--552.
\item{[UC]} D.J. Uherka and D.K. Campbell, ``The sawtooth circle map," Preprint
{\bf LA-UR} 92--
1491 (1992) (Unpublished).
\item{[UTGC]} D.J. Uherka, C. Tresser, R. Galeeva, and D.K. Campbell,
``Solvable models for the quasi-periodic transition to chaos," {\sl Phys. Lett.
A \bf 170} (1992) 189--194.
\item{[VdPVdM]} B. Van der Pol and J. Van der Mark, ``The heartbeat
considered a a relaxation oscillation and an electrical model
of the heart,'' {\it Phil. Mag.} {\bf 6}, 763-775 (1968).
\item{[Ve1]} J.J.P. Veerman, ``Symbolic dynamics and rotation numbers," {\sl
Physica \bf 134A} (1986)
543--576.
\item{[Ve2]} J.J.P. Veerman, ``Symbolic dynamics of order-preserving
orbits," {\sl Physica \bf 29D} (1987) 191--201.
\item{[Ve3]} J.J.P. Veerman, ``Irrational rotation numbers," {\sl
Nonlinearity \bf  2} (1989) 419--428.
\item{[VK]} E.B. Vul and K.M. Khanin, ``Homeomorphisms of the circle with
singularities of break
type," {\sl Russ. Math. Surveys \bf 45 } (1990)
229--230.
\item{[Wi]} A. T. Winfree, {\it The Timing of Biological Clocks}, (W. H.
Freeman,
New York (1987)).
\item {[YH]} W.-M. Yang and B.-L. Hao, ``How the Arnold Tongues become sausages
in a
piecewise
linear circle map,''{\sl Commun. Theor. Physics (Beijing)\bf 8} (1986) 1--15.

\filbreak

{\bf FIGURE CAPTIONS}

\bigskip

Fig. 1:  Typical tip and plateau maps: (a) tip or
plateau map below the critical line, (b) tip or plateau map on
the critical line, (c) tip map above the critical line,
(d) plateau map above the critical line.

\medskip
Fig. 2:  The 3-dimensional space
${\cal P}_t$ for the tip family parameters $(\Omega,S,\delta)$.

\medskip
Fig. 3:  The 3-dimensional space ${\cal P}_p$
for the plateau family parameters $(\Omega,U,\delta)$.

\medskip
Fig. 4:  Reinterpretation of certain critical maps.

\medskip
Fig. 5:  A non-monotonic left tongue boundary (of
$A_{1/20}$, above the critical line, where $\delta=0.1$
for the tip family).

\medskip
Fig. 6a:  A portion of the $q^{th}$ iterate, $f^q$, of a tip
map, at the top of $L^+_{p/q}$.

\medskip
Fig. 6b:  A portion of the  $q^{th}$ iterate, $f^q$ of a plateau map,
at the top of $L^+_{p/q}$.

\medskip
Fig. 7:  The region of circular regularity for the tip
family, for $\delta = 1/4$.  ($0 \leq p/q \leq 1/2$ with
$q \leq 11$.)

\medskip
Fig. 8:  The surface $\Pi_{S^q}$ in ${\bf C}_{S^q}$.  (The
shaded triangle is a portion of $\Pi_{S^q}$.)

\medskip
Fig. 9:  The region of zero topological entropy for the tip
family, for $\delta = 1/4$.  ($0 \leq p/q \leq 1/2$ with
$q \leq 11$.)

\medskip
Fig. 10:  Numerically generated plateau $p/q$-stability
regions, $A_{p/q}^{\oplus,{stab}}$, for $0 \leq p/q
\leq 1/2$, $q \leq 5$.

\medskip

Fig. 11:  Construction of $f'$ in the proof of Theorem 14,
for the case $m=1$.

\medskip
Fig. 12:  Stable periodic orbits must attract a turning point.

\medskip
Fig. 13:  $A_{p/q}$ and $A_{p/q}^{stab}$
for $q \leq 5$, $\delta = 1/4$.

\medskip
Fig. 14a:  $A_{p/q}^{stab}$ for $q \leq 11$, $\delta = 1/2$.
The portion of this figure above the critical line $S=S_c$
is $R_{1/2}$.

\medskip
Fig. 14b:  $A_{p/q}^{stab}$ for $q \leq 11$, $\delta = 3/4$.
The portion of this figure above the critical line $S=S_c$
is $R_{3/4}$.

\medskip
Fig. 15a:  The neighbors of $A_{0/1}^{stab}$ unzipping
from $A_{0/1}^{stab}$ at $\delta=\delta_1=1/2$.

\medskip
Fig. 15b:  The neighbors of $A_{1/2}^{stab}$ unzipping
from $A_{1/2}^{stab}$ at $\delta=\delta_2=1-2^{-1/2}$.

\medskip
Fig. 15c:  The neighbors of $A_{1/3}^{stab}$ unzipping
from $A_{1/3}^{stab}$ at $\delta=\delta_3=1-2^{-1/3}$.

\medskip
Fig. 16:  A circularly ordered chaotic 1/3-attractor and
a coexisting stable 1/2-cycle for a tip map.
$[(\Omega,S) = (0.4,2.2)$ and $\delta = 1/2.]$

\medskip
Fig. 17:  The restriction of $t^{q(i)}_{\Omega,S;\delta}$ to
a $J_k$, in the proof of Theorem 18.

\medskip
Fig. 18:  The region ${\bf R'}_{\delta}$ for $\delta =
  \delta'_3 \approx 0.1780$ is represented by the portion
of this figure above the critical line $S=S_c$.  It shows
the unzipping of the extended 1/3-region from its neighbors.
Its bigger neighbors unzip last.

\medskip
Fig. 19:  The structure of the parameter space for the tip
maps around $L_{1/2}$ with $\delta>1/2$, as shown by the graphs
of the second iterate of a lift of $t_{\Omega,S,\delta}$.


\bigskip
{\bf $\cdots$ Appendix Figures:}

\medskip
Fig. A1:  Typical circle maps of degrees -1,0,1 and 2 along
with their lifts.

\medskip
Fig. A2:  Monotone upper and lower bounds $F^+$ and
$F^-$ for a lift $F$.

\medskip
Fig. A3:  Constructing a ``stunted'' family on the interval.

\medskip
Fig. A4:  Constructing a ``stunted'' family on the circle.

\medskip
Fig. A5:  Typical graphs of:  ${\alpha}$)  $C_{S,\mu;a,b}$
and ${\beta}$)  the corresponding $(a,b)$ parameter space.

\medskip
Fig. A6:  Successive approximations to the boundary of
positive topological entropy in a stunted (also called
cutting) family.

\medskip
Fig. A7:  Parameter space for the 3-dimensional stunted
family $c_{S,\mu;a,b}$ for fixed $S$.

\bigskip
Fig. C1:  A 5-cycle for $t_{\Omega,S;1/2}$ as $(\Omega,S)$
descends along the left boundary of $A_{2/5}$.

\medskip
Fig. C2:  A lift $F$ of $t_{\Omega,S;\delta}$, where
    $\Omega = 0$.

\medskip \noindent

\medskip
Figures C3 to C15:  Tip family stability regions
$A^{stab}_{p/q}$ for $q \leq 7$ and for various $\delta$.
($\delta = $ .1, .2, .3, .4, .5, .6, .7, .8, .9, .95, .97, .99,
.9999)

\medskip
Fig. C16:  Two rational curves that intersect all $A_{p/q}$
below the critical line at nodes only.  ($\delta=0.6$ and
$q \leq 11$)

\medskip
Fig. C17:  Tip family regions $A^{stab}_{p/q}$ generated by the
Monte Carlo algorithm, with $\delta = 0.5$ and $q \leq 7$.
(Compare with the exact regions shown in Figure C7.)

\medskip
Fig. C18:  Schematic view of plateau regions above the
critical line.

\medskip
Fig. C19:  Boundary of topological chaos inside $A_{p/q}$
for the plateau family.

\medskip
Fig. C20:  Numerically generated sine family $p/q$-stability
regions for $q \leq 5$.  (Compare with Figure 10 for the
plateau family.)

\bigskip
Fig. D1:  Measure {\it versus} $S$ below the critical line,
for $\delta = 1/4, 1/2, 3/4$.
%

\end